\documentclass[journal=jacsat,manuscript=article]{achemso}

\usepackage[version=3]{mhchem} 
\usepackage{xcolor}
\usepackage{graphicx}
\usepackage{hyperref}
\usepackage{bm}
\usepackage{fancyvrb}
\usepackage{lipsum}
\usepackage{subcaption}
\usepackage{multicol}
\usepackage{float}

\author{Joel Creutzberg} 
\affiliation[Lund University]
{Division of Theoretical Chemistry, Lund University, SE- 223 62 Lund, Sweden}
\author{Erik. D. Hedeg{\aa}rd}
\affiliation[University of Southern Denmark]
{Department of Physics, Chemistry and Pharmacy, University of Southern Denmark, Campusvej 55, 5230 Odense, Denmark}
\alsoaffiliation[Lund University]
{Division of Theoretical Chemistry, Lund University, SE- 223 62 Lund, Sweden}
\email{erdh@sdu.dk, joel.creutzberg@teokem.lu.se}

\title[PE-CPP]{Polarizable embedding complex polarization propagator in four- and two-component frameworks}

\abbreviations{IR,NMR,UV}

\begin{document}

\makeatletter
\setlength\acs@tocentry@height{5.5cm}
\setlength\acs@tocentry@width{7.25cm}
\makeatother

\begin{tocentry}

\includegraphics[width=1.0\textwidth]{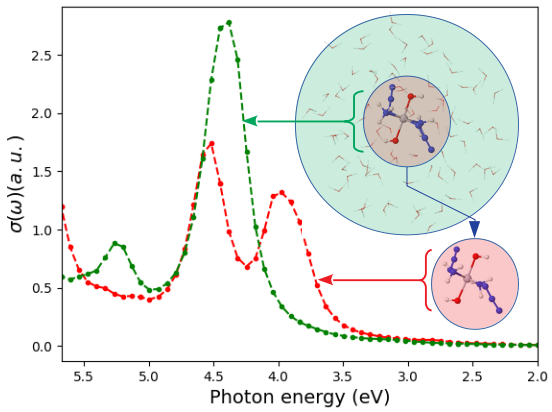}

\end{tocentry}

\begin{abstract}
Explicit embedding methods combined with the complex polarization propagator (CPP) enable modeling of spectroscopy for increasingly complex systems with a high density of states. We present the first derivation and implementation of the CPP in four- and exact two-component (X2C) polarizable embedding (PE) frameworks. We denote the developed methods PE-4c-CPP and PE-X2C-CPP, respectively. We illustrate the methods by estimating the solvent effect on UV-vis and X-ray atomic absorption (XAS) spectra of  \ce{[Rh(H2O)6]^{3+}} and \ce{[Ir(H2O)6]^{3+}} immersed in aqueous solution. We moreover estimate solvent effects on UV-vis spectra of a platinum complex that can be photo-chemically activated (in water) to kill cancer cells. Our results clearly show that inclusion of the environment is required: UV-vis and (to a lesser degree) XAS spectra can become qualitatively different from vacuum calculations. Comparison of PE-4c-CPP and PE-X2C-CPP methods shows that X2C essentially reproduces the solvent effect obtained with the 4c methods.
\end{abstract}

\section{Introduction}

Linear response functions\cite{olsen1985,helgaker2012} or \textit{polarization propagators} constitute the foundation for a large part of quantum chemical calculations on electronic excitation spectra done today. The theory has been developed to encompass many different wave function parameterizations, but its' formulation within density functional theory (DFT) is probably the most popular.\cite{runge1984,casida1995,bauernschmitt1996} While still widely employed, the original formulations suffer from several shortcomings. One shortcoming is the use of inherently non-relativistic frameworks, which is insufficient for heavier elements\cite{pyykko2012}, particularly for spectroscopies involving excitation of core-electrons\cite{norman2018}. Another shortcoming is the divergence for near-resonant perturbations, leading to numerical instabilities in spectral regions close to the electronic excitations. 

The first shortcoming has been addressed by the development of relativistic TD-DFT\cite{liu2018,saue2020,repisky2020} ranging from four-component (4c) Dirac--Kohn--Sham\cite{gao2004,gao2005,bast2009a} to various two-component frameworks\cite{wang2005,peng2005,bast2009a,kuhn2013,egidi2016}.  In parallel developments, the second shortcoming has also been addressed: it has been shown that the divergence can be removed by introducing a (phenomenological) finite lifetime of the excited states.\cite{norman2001,norman2004,kauczor2014} The term including the life-time is complex, leading to complex and real parts of the resulting response functions; hence the name \textit{complex polarization propagator} (CPP). Depending on the perturbation operator, the real and complex parts of the response functions are associated with different physical (scattering and absorption) phenomena. For calculation of electronic absorption, the CPP model exploits the imaginary part of the polarizability tensor, which is proportional to the absorption cross-section. Calculating the  absorption cross-section at a given frequency allows targeting any window of frequencies, which is another advantage of the CPP model: conventional response theory employs a bottom-up approach, calculating lower excitation energies first. This approach restricts conventional response solvers to low-lying electronic excitations and regions with low density of states. Accordingly, the CPP model allows modeling of all parts of the spectrum on the same footings and is therefore often used for core-excitation spectra.\cite{norman2018} 

Seeing that relativistic effects are expected to be large for core-electrons, it is not surprising that the CPP model has been extended to a relativistic framework\cite{villaume2010,konecny2019} where applications for both K-edge and L-edge spectra \cite{fransson2016,south2016} have been reported. Particular for the L-edge involving the 2p-orbitals, a relativistic description is pertinent since spin-orbit coupling splits the 2p-orbitals into p$_{\frac{1}{2}}$ and p$_{\frac{3}{2}}$, giving rise to  L$_{\text{II}}$ and L$_\text{III}$ edges. Generally, the inclusion of spin-orbit coupling will also entail the inclusion of a (non-degenerate) triplet manifold, resulting in a high density of states\cite{villaume2010,creutzberg2020}, giving the CPP model another advantage in relativistic calculations.   

With the development of (non-)relativistic TD-DFT in resonant as well as for non-resonant spectral regions, we seem to command a broad range of computational methods for modeling spectroscopy. Still, spectroscopic methods are used on ever more complex systems, often in the condensed phase, where interactions with a (potentially heterogeneous) environment may become important. Since large cluster models quickly become computationally intractable, efficient models to describe interactions with solvent or other environment molecules are required. One strategy is to include the environment as a structure-less continuum\cite{tomasi2005,remigio2015}. Another strategy has been to explicitly include the environment by combining a QM method with fast molecular mechanics force fields in a QM/MM hybrid.\cite{warshel1976,senn2009} We also mention another explicit embedding model employing frozen QM densities for the environment\cite{wesolowski1993,gomes2008,hofener2012}, but our paper focus on classical treatment of the environment. 

The explicit treatment of the environment has some advantages, particularly for strong interactions (e.g. hydrogen bonds) or heterogeneous environments.
Most QM/MM models are non-relativistic and employ a rather simplistic point-charge description for the environment, allowing only QM parts to be polarized by the MM part, but not vice versa. These treatments may be sufficient in some cases, e.g., for optimizing structures. However, molecular properties and excitation energies (as we focus on here) are known to be more sensitive, requiring an more advanced description of the environment electrostatics\cite{bondanza2020}.

The description of the electrostatic interactions between QM system and environment can be improved by employing a polarizable environment.\cite{warshel1976} 
Development of such polarizable embedding (PE) models dates some years back\cite{thompson1995,thompson1996}  and is still ongoing.\cite{jensen2003a,jensen2003b,gordon2007,soderhjelm2009,dziedzic2016,hedegaard2016,steinmann2019,bondanza2020} A number of works have appeared with PE models combined with linear response theory \cite{kongsted2003,sneskov2011,hedegaard2013a,list2013,hedegaard2015a,steinmann2019}  including TD-DFT.\cite{jensen2003a,yoo2007,olsen2010a,loco2016,bondanza2020} More recently, these models have also included finite life- times\cite{pedersen2014,norby2017a}, but only within a non-relativistic framework. In fact, PE models were only fairly recently extended to relativistic Hamiltonians.\cite{hedegaard2017,krause2016b}  

In this paper, we continue our development of relativistic PE models for spectroscopy: building on recent work with a relativistic linear response PE model\cite{hedegaard2017} we here report the derivation and implementation of a new relativistic PE-CPP model based on four- or exact-two-component (X2C) DFT. The environment is described by multipole moments and site-polarizabilities\cite{olsen2010a,steinmann2019} and we denote our method PE-4c-CPP or PE-X2C-CPP. The methods are implemented in the relativistic electronic structure program DIRAC\cite{saue2020, DIRAC21} and we here use them to calculate UV-vis and X-ray atomic absorption (XAS) spectra of \ce{[Rh(H2O)6]^{3+}} and \ce{[Ir(H2O)6]^{3+}} in vacuum and immersed in aqueous solution. The two systems are mainly chosen as toy-models of heavy metals in aqueous environments, where we see many future applications of the method. We also apply our newly developed method on the platinum complex \textit{trans}-\textit{trans}-\textit{trans}-\ce{[Pt(N3)2(OH)2(NH3)2]}, which is a  prototype complex for photo-activated cancer therapy.\cite{mackay2006,bednarski2006}. Therefore, its'  UV-vis spectrum has been studied extensively, although only recently by relativistic methods\cite{creutzberg2020,freitag2021} and without explicit solvent effects.

\section{Theory}

In this section, we introduce the PE model and its' extension to a CPP model in a relativistic framework. We exclusively work within the Born--Oppenheimer approximation in a second-quantization\cite{helgaker2004} Dirac--Coulomb Kohn--Sham (DC--KS) framework, using Hartree units. The method is also implemented with a  Hartree--Fock (HF) wave function. The derived expressions are generic for non-relativistic, scalar relativistic (spin-free), X2C- and 4c-KS frameworks, so differences between non-relativistic and relativistic frameworks will be highlighted. We use $p,q,r,s$ indices for general orbitals, $i,j,k,l$ for occupied orbitals and $a,b,c,d$ for unoccupied orbitals.

Relativistic PE models are rare in the literature and we accordingly have kept some detail concerning this model. A derivation of the 4c-PE model (including conventional linear response theory) was given in Ref.~\citenum{hedegaard2017}. We will often refer to this or the to literature on vacuum  CPP methods\cite{norman2005,villaume2010} to keep focus on the novel parts of the relativistic PE-4c-CPP and PE-X2C-CPP models.

\subsection*{The relativistic polarizable embedding method}

The PE model works with an energy expression similar to QM/MM models 
\begin{align}
E = E_{\mathrm{QM}} + E_{\mathrm{es}} + E_{\mathrm{ind}} + E_{\mathrm{env}} . 
\label{eq:total_energy}
\end{align}
The terms $E_{\mathrm{QM}}$ and $E_{\mathrm{env}}$  describe interactions exclusively within the QM and environment parts, respectively. Since $E_{\mathrm{env}}$ does not depend on any wave function parameters, it will not figure in the response equations in the next section and we will not consider this term further here. The terms $E_{\mathrm{es}}$ and $E_{\mathrm{ind}}$ describe the interaction between QM and environment: the former is the electrostatic interactions and the latter the mutual polarization of the environment and QM parts (this term is missing from most QM/MM models). We discuss  $E_{\mathrm{QM}}$, $E_{\mathrm{es}}$ and $E_{\mathrm{ind}}$ from Eq.~\eqref{eq:total_energy} in more detail  below.

The QM energy in a 4c-KS framework based on a Dirac--Coulomb Hamiltonian  is given\cite{saue2002,salek2002}  
\begin{align}
E_{\mathrm{QM}} = \sum_{pq}h_{pq}D_{pq} + \sum_{pq}j_{pq} D_{pq} + E_{\mathrm{xc}}[\rho] + E_{\mathrm{nn}} ,  
\label{eq:energy_func}
\end{align}
where $D_{pq} = \langle 0 \vert \hat{a}^{\dagger}_p \hat{a}_q\vert 0\rangle$ is an element of the one-electron reduced density matrix with creation ($\hat{a}^{\dagger}_p$) and annihilation ($\hat{a}_q $) operators working on orbitals $p$ and $q$, respectively. The KS reference determinant is denoted as $\vert 0 \rangle$. Further,  $\rho$ is the electron density and $h_{pq}$ are integrals over the one-electron operator, which 
for a 4c framework reads
\begin{align}
	\hat{h} = \hat{h}_{D} + \hat{V}_{\mathrm{ext}} =  
	\left(\begin{array}{cc} \bm{0}_2 & c(\bm{\sigma}\cdot \hat{\mathbf{p}}) \\
		c(\bm{\sigma}\cdot \hat{\mathbf{p}})   &   - 2c^{2} \mathbf{I}_2
	\end{array}\right) + 
	\left(\begin{array}{cc}  V_{\mathrm{ext}}\mathbf{I}_2  & \bm{0}_2 \\
		\bm{0}_2   &  V_{\mathrm{ext}}\mathbf{I}_2
	\end{array}\right) . 
	\label{eq:h_D}
\end{align}
For the remaining terms in Eq.~\eqref{eq:energy_func} and the definition of constants and operators in Eq.~\eqref{eq:h_D}, we refer to the definitions in  Refs.~\citenum{salek2002,saue2002}. We define $E_{\mathrm{xc}}[\rho]$ including exact exchange as in Ref.~\citenum{saue2002}. An analytic solution can be obtained for the one-electron part of  Eq. \eqref{eq:energy_func} and we denote this wave function 
\begin{align}
	\label{eq:4c-wavefunc}
\vert \psi^{4c} \rangle =  \left(\begin{array}{c} \vert \psi^{L} \rangle  \\ \vert \psi^{S} \rangle\end{array} \right), 
\end{align}
where $\vert \psi_L\rangle$ and $\vert \psi_S \rangle$ are the large- and small-component wave functions, respectively. These are themselves two-component wave functions, yielding the 4c form of $\vert \psi^{4c} \rangle$ as required by the structure of Eq.~\eqref{eq:h_D}. We will use this nomenclature when we discuss X2C methods below, but for now we focus on the 4c-KS method.  
   
The optimal energy in Eq.~\eqref{eq:energy_func} is obtained through diagonalization of the KS-matrix, $\mathbf{F}$, which is equivalent to  requiring the off-diagonal elements to vanish
\begin{align}
F_{ia} =  \langle i \vert \hat{f}_0 \vert a\rangle = 0.  \label{KS_matrix_elements}
\end{align}
We have here defined the KS-operator in second-quantization 
\begin{align}
    \hat{f}_0 = \sum_{pq}f_{pq} \hat{a}^{\dagger}_{p}a_{q}   .
\label{eq:ks_operator}
    \end{align}
Expressions for the integrals in  $f_{pq}$ can be found in the literature (see, e.g., Ref.~\citenum{saue2002}); they contain the one electron part, i.e., Eq.~\eqref{eq:h_D}, and the Coulomb and exchange--correlation terms. Differences to the non-relativistic theory manifest in the $4\times 4$ matrix form of the operator in Eq.~\eqref{eq:h_D}; the KS reference state, $\vert 0 \rangle$, is accordingly on a similar form as Eq.~\eqref{eq:4c-wavefunc}. In a finite-orbital basis, the individual orbitals can be written 
\begin{align}
	\label{eq:4-spinor}
\vert p\rangle =  \sum^{N_L}_{\kappa} \left(\begin{array}{c} \vert \chi^{L}_{\kappa}\rangle  \\ 0\end{array} \right) C^{L}_{\kappa} 
+ \sum^{N_s}_{\lambda} \left(\begin{array}{c} 0 \\ \vert \chi^{S}_{\lambda}\rangle \end{array}\right)   C^{S}_{\lambda} && ; &&  
\vert \chi^{X}_{i} \rangle = \left(\begin{array}{c}  \vert\chi^{X}_{i\alpha}\rangle \\ \vert \chi^{X}_{i\beta}\rangle \end{array}\right) ,
\end{align}
where the primitive basis functions $\vert \chi^{L}_{\kappa} \rangle$ and $\vert \chi^{S}_{\lambda} \rangle$, similar to Eq.~\eqref{eq:4c-wavefunc}, are two-component spinors $\vert \chi^{X}_{i}\rangle$ ($X = L,S$) with  $\alpha$- and $\beta$ parts. The expansion coefficients  $C^{L}_{\kappa}$ and $C^{S}_{\lambda}$ for the large and small component spinors are in the {\sc{DIRAC}} program partitioned according to a quaternion symmetry scheme, reducing to complex or real algebra, depending on the spatial symmetry group. However, since the implementation of the PE model currently does not exploit spatial symmetry, we will always use full quaternion symmetry. 

From Eq.~\eqref{eq:4-spinor} it follows that the matrix representations of the one-electron Hamiltonian and the KS matrix in Eq.~\eqref{KS_matrix_elements} have the following structures 
\begin{align}
\label{eq:4-4-fock-matrix}
\mathbf{h}_D = \left( \begin{array}{cc} \mathbf{h}^{LL}_D & \mathbf{h}^{LS}_D \\
	\mathbf{h}^{SL}_D & \mathbf{h}^{SS}_D   \end{array} \right)  && 
\mathbf{F} = \left( \begin{array}{cc} \mathbf{F}^{LL} & \mathbf{F}^{LS} \\
\mathbf{F}^{SL} & \mathbf{F}^{SS}   \end{array} \right) , 
\end{align}  
where each element is a $2\times 2$ block with the generic matrix elements  $h^{XY}_{D,\kappa\lambda} = \langle\chi^{X}_{\kappa} \vert \hat{h}_D \vert \chi^{Y}_{\lambda}\rangle$ and $F^{XY}_{\kappa\lambda} = \langle\chi^{X}_{\kappa} \vert \hat{f}^{\mathrm{KS}}_0 \vert \chi^{Y}_{\lambda}\rangle$. 

We  now turn to the PE operators giving rise to the two terms $E_{\mathrm{es}}$ and $E_{\mathrm{ind}}$ in Eq.~\eqref{eq:total_energy}. These operators have been described  earlier\cite{olsen2010b,hedegaard2013a,pedersen2014,steinmann2019} with few exceptions\cite{krause2016a,hedegaard2017} within non-relativistic frameworks. The electrostatic energy contribution is obtained as the expectation value of $\hat{V}^{\mathrm{es}}$
\begin{align}
	E_{\mathrm{es}} = \langle 0 \vert \hat{V}^{\mathrm{es}}\vert 0 \rangle,  
	\label{eq:Ees}
\end{align}
where the operator $\hat{V}$  is defined  
\begin{align}
	\label{eq:V_es}
	\hat{V}^{\mathrm{es}} = \sum_s \sum_{|k|=0} \frac{(-1)^{|k|}}{k!} \Bigr( \mathbf{M}_s^{(k)} \sum_{pq} \mathbf{t}^{(k)}_{pq,s} \hat{a}^{\dagger}_{p}\hat{a}_{q} + \mathbf{M}_s^{(k)} \sum_{M} Z_M  \mathbf{T}_{Ms}^{(k)}   \Bigr) . 
\end{align}
We have in accordance with previous literature\cite{olsen2010b,olsen2011,hedegaard2013a,pedersen2014,hedegaard2017,steinmann2019} used a three-dimensional multi-index $k=(k_x , k_y, k_z)$ for multipoles $\mathbf{M}^{(k)}_s$ located at sites $s$; thus $k=0$ corresponds to point charges, $\mathbf{M}^{(0)}_s = q_s$, $k=1$ corresponds to components of the dipole vector, $\mathbf{M}^{(1)}_s = \bm{\mu}_s =(\mu_{s,x},\mu_{s,y},\mu_{s,z})$, and so on. This notation is also used in the definition of the integrals 
$\mathbf{t}^{(k)}_{pq,s} = -\langle p \vert \mathbf{T}^{(k)}_{s} \vert q\rangle $ over interaction operators, $\mathbf{T}^{(k)}_{j}$, defined as in the monograph by Stones.\cite{stone2002} We construct the operator in Eq.~\eqref{eq:V_es} employing the library {\sc{PELib}}\cite{pelib} which previously has been interfaced with {\sc{DIRAC}}.\cite{hedegaard2017} The multipole moments are obtained as  atom-localized multipole moments from the LoProp method\cite{gagliardi2004}. Since this requires a separate QM calculation, the computational cost is kept low by dividing the environment into fragments. Multipole moments are in this way calculated for each fragment. 

The term accounting for the mutual polarization of QM and environment regions in Eq.~\eqref{eq:total_energy} is given 
\begin{equation}
	E_{\mathrm{ind}}=-\frac{1}{2}(\bm{\mu}^{\mathrm{ind}})^T  \langle 0 \vert \hat{\bm{\mathcal{E}}}\vert 0 \rangle 
	\label{eq:Eind},
\end{equation}
where the field vector operator, $\hat{\bm{\mathcal{E}}}$, is arranged as a vector where each  entry contains the field operator for a \textit{polarizable} site, $\hat{\bm{\mathcal{E}}} =(\hat{\bm{\mathcal{E}}}_1,\ldots,\hat{\bm{\mathcal{E}}}_s ,\ldots,\hat{\bm{\mathcal{E}}}_S)$. The induced dipole moment is a vector arranged similarly. The field on site $s$ has contributions from QM electrons and nuclei as well as the multipoles, leaving the total field operator for site $s$ as
\begin{align}
	\label{eq:total_field}
	\hat{\bm{\mathcal{E}}}_s = \hat{\bm{\mathcal{E}}^{\mathrm{e}}}_s + \bm{\mathcal{E}}^{\mathrm{nuc}}_s +  \bm{\mathcal{E}}^{\mathrm{es}}_s  .
\end{align} 
The  electronic and nuclear fields in Eq.~\eqref{eq:total_field} are given as 
\begin{align}
	\hat{\bm{\mathcal{E}}}^e_{s} &= \sum_{pq}\mathbf{t}^{(1)}_{pq,s}\hat{a}^{\dagger}_{p}\hat{a}_{q} =  -\sum_{pq} \bigr \langle p \vert \frac{\mathbf{r}-\mathbf{r}_{s}}{| \mathbf{r}-\mathbf{r}_{s} |^3} \vert q  \bigl \rangle \hat{a}^{\dagger}_{p}\hat{a}_{q}   \\    
	\bm{\mathcal{E}}_{s}^n &= -\sum_{M} Z_{M}\mathbf{T}^{(1)}_{Ms} = \sum_M \frac{Z_M(\mathbf{R}_M - \mathbf{r}_{s})}{|\mathbf{R}_M - \mathbf{r}_{s}|^3} . 
\end{align}
The individual induced dipole moments $\bm{\mu}^{\mathrm{ind}}$ in Eq.~\eqref{eq:Eind} are obtained by self-consistently solving the matrix equation\cite{applequist1972} $ \bm{\mu}^{\mathrm{ind}} = \mathbf{R}^{Relay}\bm{\mathcal{E}}$ in each self-consistent-field iteration. The matrix $\mathbf{R}^{Relay}$ is given explicitly in previous publications.\cite{olsen2010b,olsen2011,hedegaard2013a,pedersen2014,hedegaard2017,steinmann2019} Here it is sufficient to note that it does not contain any wave function parameters and the diagonal contains polarizabilities (isotropic or an-isotropic) for each polarizable site. The polarizabilities are  obtained with the LoProp method\cite{gagliardi2004}, employing the same fragments as for the multipole moments.  

From the expressions $E_{\mathrm{es}}$ and $E_{\mathrm{ind}}$ in Eqs.~\eqref{eq:Ees} and \eqref{eq:Eind}, we have previously shown\cite{hedegaard2017} that the KS-operator due to a polarizable environment, is modified with the PE-operator, $\hat{v}^{\mathrm{PE}}$: 
\begin{align}
	\hat{f}^{\mathrm{tot}} & = \hat{f}_0 + \hat{v}^{\mathrm{PE}} \label{eq:fock_pe-1} \\
	 \hat{v}^{\mathrm{PE}} &= \hat{V}^{\mathrm{es}} - \bm{\mu}^{\mathrm{ind}} \hat{\bm{\mathcal{E}}}^{\mathrm{e}} =  \hat{V}^{\mathrm{es}} - \langle 0 \vert \hat{\bm{\mathcal{E}}} \vert 0 \rangle ^T\, \mathbf{R}^{Relay}\, \hat{\bm{\mathcal{E}}}^{\mathrm{e}} \label{eq:fock_pe-2} .
\end{align}
Contrary to a non-relativistic framework, the $\hat{v}^{\mathrm{PE}}$ operator will attain a $4\times 4$ matrix structure. The only non-zero matrix elements of $\hat{v}^{\mathrm{PE}}$ over four-component spinors (Eq.~\ref{eq:4-spinor}) are within the large-large or small-small blocks.\cite{hedegaard2017} Thus, only $\mathbf{F}^{LL}$ and $\mathbf{F}^{SS}$ blocks are modified in matrix representation of the total KS operator 
\begin{align}
\mathbf{F}^{\mathrm{KS}} = \left( \begin{array}{cc} \mathbf{F}^{LL} +  \mathbf{v}^{\mathrm{PE},LL}   & \mathbf{F}^{LS} \\
\mathbf{F}^{SL} & \mathbf{F}^{SS} + \mathbf{v}^{\mathrm{PE},SS}  \end{array} \right). 
\end{align}
where $\mathbf{v}^{\mathrm{PE},XY}$ contains the elements,  $\hat{v}^{\text{PE},XY}_{\kappa\lambda}=\langle\chi^{X}_{\kappa}\vert \hat{v}^{\text{PE}}\vert \chi^{Y}_{\lambda}\rangle$. 

In addition to PE-4c-KS (and HF), our implementation also covers an X2C Hamiltonian. In X2C methods, the large- and small-component equations resulting from the form of $\vert \psi^{4c}\rangle$ in Eq.~\eqref{eq:4c-wavefunc} are decoupled. Formally, this is done by the decoupling matrix $\mathbf{U}$
\begin{align}
	\label{x2c-U}
	\mathbf{U}^{\dagger}\mathbf{h}_D\mathbf{U} = \left( \begin{array}{cc} \mathbf{h}_{++} & 0  \\
		0 & \mathbf{h}_{--}   \end{array} \right) && \mathbf{U} = \left( \begin{array}{cc} \mathbf{\Omega}_{+} & -\mathbf{R}^{\dagger}\mathbf{\Omega}_{-}  \\
		\mathbf{R}\mathbf{\Omega}_{+} & \mathbf{\Omega}_{-}   \end{array} \right),  
\end{align}	
where $\mathbf{\Omega}_{+} = \left(1 + \mathbf{R}^{\dagger}\mathbf{R}  \right)^{-\frac{1}{2}}$, $\mathbf{\Omega}_{-} = \left(1 + \mathbf{R}\mathbf{R}^{\dagger}  \right)^{-\frac{1}{2}}$, and $\mathbf{R}$ describes the  coupling of the large and small component wave functions\cite{saue2011}. The decoupling in  Eq.~\eqref{x2c-U} allows us to focus only on the positive energy solutions 
\begin{align}
	\mathbf{h}_{++}\vert \psi^{2c}_{+} \rangle = E \vert \psi^{2c}_{+} \rangle, 
\end{align}	
where $\vert \psi^{2c}_{+} \rangle = \left(1 + \mathbf{R}^{\dagger}\mathbf{R}  \right)^{\frac{1}{2}} \vert \psi^{L} \rangle$ can be expanded in the large-component basis functions, thereby reducing the computational cost significantly. For HF or DFT methods, a transformation of the two-electron integrals in the the KS matrix (Eq.~\ref{KS_matrix_elements}) is in principle required, but this transformation would make the calculation more expensive than the corresponding 4c calculation. In the PE-X2C implementation, we follow the general procedure in DIRAC where $\mathbf{U}$ is constructed from the one-electron Hamiltonian and two-electron integrals from the large-component basis are used unmodified\cite{saue2011}. Leaving the two-electron part untransformed corresponds to neglecting two-electron spin-spin and spin-orbit contributions, but these can be very accurately approximated by atomic mean-field corrections\cite{hess1996} (see the Computational Details Section).
 
The extraction of $\mathbf{U}$ and construction of the two-component (X2C) wave function in DIRAC occur prior to the (self-consistent) calculations of the PE contributions in Eq.~\eqref{eq:fock_pe-2}. Hence, the PE contributions in PE-X2C are always constructed from the one-electron reduced density matrix corresponding to the X2C reference state. We verify that this works well by comparison of PE-X2C-CPP and PE-4c-CPP in the Results and Discussion Section. An alternative route is to use a so-called mean-field X2C strategy\cite{sikkema2009}, i.e., solve the PE-4c  KS (or HF) equations and perform the transformation on the final KS (or Fock) matrix. We leave an investigation of this route for future studies.

We will in the next section provide the theory behind the PE-CPP method, but without distinguishing between 4c and X2C frameworks, since their equation structures are equivalent. However, we note that the transformation in Eq.~\eqref{x2c-U} should be carried over to the property operators to avoid so-called picture change errors \cite{Baerends_1990,Kell1998PictureCA,Dyall2000}. Following the general implementation in the DIRAC program, all property operators are always transformed.

\subsection*{The relativistic polarizable embedding complex polarization propagator}

Introducing $ \hat{v}^{\text{PE}}$ in Eq.~\eqref{eq:fock_pe-1} will also lead to modified response equations due to the polarizable environment \cite{jensen2003a,yoo2007,olsen2010a,loco2016,bondanza2020,pedersen2014,norby2017a}. We now examine these changes within a relativistic PE model that includes a  finite life time. 

In response theory we add a time-dependent interaction operator, $\hat{V}(t)$, as perturbation to the Hamiltonian; the operator defines a periodic, time-dependent field   
\begin{equation}
	\label{eq:period_perturb}
	\hat{V}(t)= \sum_{k} e^{-i\omega_k t} \sum_X \mathcal{E}_X (\omega_k) \hat{X}, 
\end{equation} 
where $\hat{X}$ is a property operator, $\mathcal{E}_X (\omega_k)$ are field-strength parameters (defined as Fourier components) with frequencies $\omega_k$. We follow the nomenclature from Ref.~\citenum{hedegaard2017}, labeling fields from perturbation operators  with frequencies, $\omega_k$, while components of
the fields from QM region and  multipoles at the
polarizable sites in the environment are labeled as  in
the previous section (cf.~Eq.~\ref{eq:total_field}).

We will here work with a time-dependent KS reference determinant, parameterized as 
\begin{equation}
\label{eq:param_wavefunction}
| \tilde{0}(t) \rangle= e^{-\hat{\kappa}(t)}| 0 \rangle , 
\end{equation}
where $\kappa(t)$ is the time-dependent orbital-rotation operator. Since we are here only interested in linear response theory, we keep only linear parameters in the perturbation expansion $\kappa(t) = \kappa^{(1)}(t) + \cdots$ which we define
\begin{equation}
\label{eq:kappa}
\hat{\kappa}^{(1)}(t)= \sum_{k}\sum_{ai}\Bigr(\kappa_{ai}(\omega_k)\hat{q}^{\dagger}_{ai}-\kappa^*_{ai}(-\omega_k)\hat{q}_{ai}\Bigl)e^{-i\omega_k t} .
\end{equation}
The parameters $\kappa_{ai}(\omega_k) = \sum_{X}\mathcal{E}_X(\omega_k) \kappa^{\omega_k}_{ai}$ are linear in the field-strength and we have  defined $\hat{q}_{ai} = \hat{a}^{\dagger}_{i} \hat{a}_a$. 
Using the quasi-energy formulation of response theory\cite{christiansen1998}, a time-dependent variational principle   
is established through the time-averaged quasi-energy  
\begin{equation}
	\label{eq:quasi_energy-0}
	\{Q(t) \}_T=\frac{1}{T}\int_{-T/2}^{T/2} \langle \tilde{0}(t) |\big( \hat{H} - i \frac{\partial}{\partial t} \big)| \tilde{0}(t) \rangle  dt , 
\end{equation}
from which response functions and response equations can be  derived\cite{christiansen1998,helgaker2012}. The  Hamiltonian is given  $\hat{H} = \hat{H}^{\mathrm{tot}} + \hat{V}(t)$: the first part, $\hat{H}^{\mathrm{tot}}$, contains the usual components of the KS-DFT  framework\cite{salek2005} corresponding to $E_{\mathrm{QM}}$ in Eq.~\eqref{eq:energy_func} and additionally operators corresponding to $E_{\mathrm{es}}$ and $E_{\mathrm{ind}}$ (Eqs.~\ref{eq:Ees}--\ref{eq:Eind})\cite{hedegaard2017}. The second part of $\hat{H}^{\mathrm{tot}}$ is the external pertubating field in Eq.~\eqref{eq:period_perturb}. By inserting $\hat{H}$ in Eq.~\eqref{eq:quasi_energy-0} we can write the quasi energy\cite{salek2005} 
\begin{align}
		\label{eq:quasi_energy-1}
	\{Q_{\mathrm{tot}}(t) \}_T = \{\bar{Q}(t) \}_T + \sum_{k}\sum_{X}\mathcal{E}_X (\omega_k) \{ \bar{Q}_X (t)\}_T , 
\end{align}
where $\{\bar{Q}(t) \}_T$ has origin in the field-independent part of $\hat{H}$ (and also includes the time-derivative of Eq.~\ref{eq:quasi_energy-0})  and $\{ \bar{Q}_X (t)\}$ contains the time-dependent part with the property operator, $\hat{X}$. Following Ref.~\citenum{hedegaard2017}, we can divide both terms according to  
\begin{align}
	\label{eq:quasi_energy-2} 
	 \{\bar{Q}(t) \}_T = Q_0 + \{Q^{\mathrm{PE}}(t) \}_T && 
	 	\{\bar{Q}_X(t) \}_T = \{Q_{X}(t) \}_T + \{Q^{\mathrm{PE}}_{X}(t) \}_T , 
\end{align}
where $\{Q^{\mathrm{PE}}(t) \}_T  $ and  $\{Q^{\mathrm{PE}}_{X}(t) \}_T$ are the additional terms, arising due to the (polarizable) environment (see Ref.~\citenum{hedegaard2017} for explicit expressions). Note also that the field-dependent term in Eq.~\eqref{eq:quasi_energy-2} has been augmented with a contribution due to the polarizable environment; this contribution does not have origin in $\hat{H}^{\mathrm{tot}}$ and was not included in initial work with the (non-relativistic) PE-model.\cite{soderhjelm2009,olsen2010a,sneskov2011,hedegaard2015a} The contribution arise as the perturbing field $\hat{V}(t)$ also  induces a field, $\tilde{\bm{\mathcal{E}}}(t)$, within the environment, modifying the total induced field in Eq.~\eqref{eq:total_field} to $\tilde{\bm{\mathcal{E}}}^{\mathrm{tot}}_s = \tilde{\bm{\mathcal{E}}}^{\mathrm{e}}_s + \bm{\mathcal{E}}^{\mathrm{nuc}}_s +  \bm{\mathcal{E}}^{\mathrm{es}}_s + \tilde{\bm{\mathcal{E}}}_{s}(t) $ within the linear response framework.  This effect is denoted the effective external field (EEF) contribution.\cite{jensen2005,list2016}

Response equations can now be obtained through derivatives of the quasi-energy. In this last part of the derivation, our paper differs  from the previous relativistic PE models.\cite{hedegaard2017,krause2016a} It has been shown for exact-state response theory\cite{kristiansen2009} that employing a wave function without reference to the excited state life-time (as the one in Eq.~\ref{eq:param_wavefunction}), corresponds to wave function whose norm is constant in time. This is a somewhat absurd situation since  the corresponding excited states thus have infinite lifetime.\cite{kristiansen2009,helgaker2012} A physically more pleasing expression including an excited state life-time can be obtained by employing a phenomenological, common life-time for all excited states ($\gamma$) in the definition of the wave function. As shown (for an exact-state)  in Refs.~\citenum{kristiansen2009,helgaker2012} this leads identical response equations, but with state-energy differences $\omega_k$ replaced by the complex entity $z = \omega_k +i\gamma $.\cite{norman2001,norman2005,kristiansen2009,villaume2010,kauczor2014} By analogy, we arrive at the PE-CPP equations by inserting the phase-isolated KS determinant in Eq.~\eqref{eq:param_wavefunction} with the orbital rotation operator in Eq.~\eqref{eq:kappa} into the quasi energy, Eqs.~\eqref{eq:quasi_energy-1}--\eqref{eq:quasi_energy-2}, while replacing $\omega\rightarrow z$.  
This  leads to the following response equations 
\begin{align}
	\label{eq:solution-vector}
\bm{\kappa}^{z} = - \Bigl(\bar{\mathbf{E}}^{[2]} - z\mathbf{S}^{[2]}\Bigr)^{-1} \bar{\mathbf{E}}^{[1]}_X .
\end{align}
The bars over the electronic Hessian, $\mathbf{\bar{E}}^{[2]}$, as well as the property gradient vector, $\mathbf{\bar{E}}^{[1]}_X$, signify that they both contain additional contributions due to the polarizable environment.  For the Hessian, we obtain   
\begin{align}
	\bar{\mathbf{E}}^{[2]} = \left(\begin{array}{cc} \mathbf{A} & \mathbf{B} \\
	                                           \mathbf{B}^* & \mathbf{A}^*   \end{array}  \right) 
\end{align} 
with 
\begin{align}
A_{ai;bj} & =	\langle  0 \vert [\hat{q}^{\dagger}_{ai},[ \hat{q}_{bj},\hat{f}^{\mathrm{KS}}_0]] \vert 0 \rangle + \langle  0 \vert [\hat{q}^{\dagger}_{ai}, \hat{v}^{\mathrm{Hxc}}] \vert 0 \rangle  \notag \\ & + \langle  0 \vert [\hat{q}^{\dagger}_{ai},[ \hat{q}_{bj}, \hat{v}^{\mathrm{PE}}]] \vert 0 \rangle + \langle  0 \vert [\hat{q}^{\dagger}_{ai}, \hat{v}^{\mathrm{ind}}] \vert 0 \rangle  \label{response-A} \\
B_{ai;bj} & =	\langle  0 \vert [\hat{q}_{ai},[ \hat{q}_{bj},\hat{f}^{\mathrm{KS}}_0]] \vert 0 \rangle + \langle  0 \vert [\hat{q}_{ai}, \hat{v}^{\mathrm{Hxc}}] \vert 0 \rangle  \notag \\ & + \langle  0 \vert [\hat{q}_{ai},[ \hat{q}_{bj}, \hat{v}^{\mathrm{PE}}]] \vert 0 \rangle + \langle  0 \vert [\hat{q}_{ai}, \hat{v}^{\mathrm{ind}}] \vert 0 \rangle \label{response-B}  .
\end{align}
The two first terms in Eqs.~\eqref{response-A}--\eqref{response-B} are the vacuum contributions, whereas the following  two  terms are the additional terms originating from the PE part. The first term contains the KS operator in Eq.~\eqref{eq:ks_operator} while the second term with the $\hat{v}^{\text{Hxc}}$ operator is the Hartree-exchange-correlation kernel (note that this term was left out in Eq.~58 of Ref.~\citenum{hedegaard2017} by mistake). In the last two terms, we identify  the PE operator in Eq.~\eqref{eq:fock_pe-2} and we have additionally defined  
\begin{align}
	\hat{v}^{\mathrm{ind}} & =  - \langle 0 \vert [\hat{q}_{bj},  \hat{\bm{\mathcal{E}}}^{\text{e}}]\vert 0\rangle^T \, \mathbf{R}^{Relay}\, \hat{\bm{\mathcal{E}}}^{\mathrm{e}}.
\end{align}	
Meanwhile, the elements of the property gradient becomes 
\begin{align}
	\bar{E}^{[1]}_{ai,X} = \langle 0 \vert [\hat{q}_{ai},\hat{X} ] \vert 0 \rangle + \frac{\text{d}\bm{\mathbf{\mu}}^{\mathrm{ind}}_{\text{ext},X}(\omega_k)}{\text{d}\bm{\mathcal{E}}_X (\omega_k)}  \langle 0 \vert [\hat{q}_{ai},\hat{\bm{\mathcal{E}}}^{\mathrm{e}} ] \vert 0 \rangle  , 
\end{align}	
where the last term, containing $\bm{\mu}^{\mathrm{ind}}_{\text{ext},X}(\omega_k) = \mathbf{R}^{Relay}\bm{\mathcal{E}}_X (\omega_k)$, gives rise to the EEF effect.  

Explicit construction of the  Hessian in Eq.~\eqref{eq:solution-vector} is computationally expensive, but is avoided by  expanding the solution vector $\bm{\kappa}^{z} = \sum_{i=1}^n \mathbf{b}_i a_{i}(z)$ in a set of trial vectors, $\{\mathbf{b}_i\}$, meaning that only  
\begin{align}
	\bm{\sigma}_i = \bar{\mathbf{E}}^{[2]}\mathbf{b}_i  ,
\end{align}
is required. Several algorithms for constructing these trial vectors have been presented (see, e.g.~Ref.~\citenum{kauczor2011} and reference therein). We rely on the scheme developed in DIRAC, where the trial vectors are divided according to their Hermitian and time-reversal symmetries.\cite{saue2003,bast2009a,villaume2010} This scheme has also been extended to response equations with complex frequencies\cite{villaume2010} but the addition of the PE terms will not lead to any changes. Thus, the only modifications of the vacuum equations  are due to the modifications in $\bar{\mathbf{E}}^{[2]}$ and $\bar{\mathbf{E}}^{[1]}_X$,  leaving $\bm{\sigma}_i$ to be   
\begin{align}
\sigma_{ai} & = \langle 0|[-\hat{q}_{ai}, \tilde{f}_0]|0\rangle + \langle 0|[-\hat{q}_{ai}, \tilde{v}^{\text{Hxc}}]|0\rangle \notag \\
& + \langle 0|[-\hat{q}_{ai}, \tilde{v}^{\text{PE}}]|0\rangle + \langle 0 |\tilde{\bm{\mathcal{E}}}^e | 0 \rangle^T \, \mathbf{R}^{Relay}\, \langle 0  |[-q_{ai},\hat{\bm{\mathcal{E}}}^e]| 0 \rangle \notag \\
& = \tilde{f}^{\text{KS}}_{0,ai} + \tilde{v}^{\text{Hxc}}_{ai} + \tilde{v}^{\text{PE}}_{ai} + (\tilde{\bm{\mu}}^{\text{e}})^T \cdot \bm{\mathcal{E}}^{\text{e}}_{ai}. 
\end{align}
Note that we have used the nomenclature $\tilde{O}$ for one-index transformed operators.\cite{olsen1985}  

\section{Computational Details \label{methods}}

\begin{figure}[htb!]
	\includegraphics[width=0.75\textwidth]{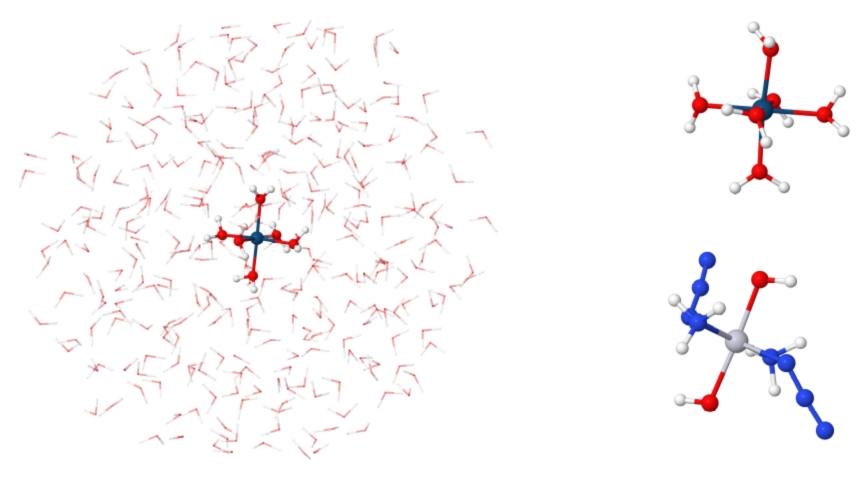}
	\caption{The solvated sytems used in the PE calcuations for \ce{[Rh(H2O)6]^{3+}} and \ce{[Ir(H2O)6]^{3+}} (shown with and without solvent, using \ce{[Ir(H2O)6]^{3+}} as example). The solvated   \textit{trans}-\textit{trans}-\textit{trans}-\ce{Pt(N3)2(OH)2(NH3)} (also shown without solvent below \ce{[Ir(H2O)6]^{3+}}) is treated similar.}
	\label{fig:iridium_in_water}
\end{figure}

\textbf{Structure optimizations of \ce{[Rh(H2O)6]^{3+}} and \ce{[Ir(H2O)6]^{3+}}:} For the solvated structures, we employed the Maestro builder\cite{qsite2021} to construct the metal ions, immersed in a cubic solvent box (25 $\times$ 25 $\times$ 25 \AA) of explicit water molecules, modeled with TIP3P\cite{jorgensen1983}. The  systems were optimized with QM/MM using QSite\cite{philipp1999,murphy2000,qsite2021}. The QM regions were comprised of the \ce{[Rh(H2O)6]^{3+}} and \ce{[Ir(H2O)6]^{3+}} ions, for which we employed the B3LYP functional\cite{Becke1988,Lee1988,Becke1993} combined with the LACVP$^*$ basis set.\cite{bs720301hdp,ditchfield1971,hariharan1973,hay1985}
The MM region was relaxed with the truncated Newton method. As an example, we show the optimized structures of \ce{[Ir(H2O)6]^{3+}} in Figure \ref{fig:iridium_in_water}. Vacuum structures where optimized with same same functional and basis set and resulted in symmetric structures ($C_i$ symmetry, although note that this was not used in the calculations).

\textbf{Structure of \textit{trans}-\textit{trans}-\textit{trans}-\ce{[Pt(N3)(OH)2(NH3)2]} in water:} Optimization of the platinum complex (denoted \textit{trans}-Pt from here) was more challenging than the simple aqua ions. We employed a similar strategy as for the aqua ions, although the box-size of explicit water molecules was larger (35 $\times$ 35 $\times$ 35 \AA). 	 

First, the structure of the \textit{trans}-Pt complex was optimized with frozen surroundings using the QM/MM method implemented in QSite\cite{philipp1999,murphy2000,qsite2021} (and same functional/basis set  as for the aqua ions). From this system, a sphere of 6 {\AA} was optimized with the  BP86\cite{Becke1988,perdew1986} functional and the def2-SV(P) basis set\cite{Andrae1990a,Schafer1992,Weigend2005a} in ORCA \cite{orca} (water molecules within 4 {\AA} of the  \textit{trans}-Pt complex was allowed  to structurally relax; the rest was kept frozen). For computational efficiency, a pre-optimization with PBEh-3c\cite{grimme2015} method was done. After this, the 6 {\AA} sphere was re-inserted into the original box and a 10 {\AA} sphere was cut out. All PE calculations were performed in this 10 {\AA} system.

\textbf{Polarizable embedding potentials:} Atom-centered multipoles (charges, dipoles, quadrupole moments) and static anisotropic polarizabilities were calculated with the LoProp\cite{gagliardi2004} method in Molcas\cite{molcas2019} on individual fragments (each water molecule constitutes a fragment). The setup of the calculations was done using the PEAS\cite{jogvan_thesis} script. The calculations employed the B3LYP functional and a A-6-31PGP basis set (note that this basis set is specially constructed to fit the LoProp method).  
 
\textbf{Complex polarization propagator calculations:} Calculation of frequency-dependent electric dipole polarizabilities were carried out on \ce{[Rh(H2O)6]^{3+}}, \ce{[Ir(H2O)6]^{3+}} and the \textit{trans}-Pt complex with CPP for three cases, namely 1) the vacuum optimized structures (denoted vacuum), 2) the structures immersed in a solvent described by a potential (denoted PE), and 3) as case 2), but with the addition of an effective external field (denoted PE-EEF). We additionally carried out a set of calculations without a PE potential, using the QM/MM optimized structures (i.e.~they are optimized within solvent). This will allow us to separate structural and electronic solvent effects.
The calculation were done in the frequency ranges corresponding to the UV-vis and various XAS regions, using CPP and PE-CPP in a development version of {\sc{DIRAC}}.\cite{saue2020,DIRAC21} For the UV-vis spectra the CPP calculations were carried out in the ranges of 1.7--8.8 eV for \ce{[Rh(H2O)6]^{3+}}, 1.8--8.8 eV for \ce{[Ir(H2O)6]^{3+}} and 2.0-6.8 eV for \textit{trans}-Pt 
with steps of 0.068 eV (0.0025 Hartree). In order to obtain more data points for the UV-vis spectra of  \ce{[Rh(H2O)6]^{3+}} and \ce{[Ir(H2O)6]^{3+}}, a second scan was performed with start and end 0.3 eV before the original scans. The core spectra for the L$_{II}$/L$_{III}$ edges were calculated in the ranges of   3104.0-- 3124.2 eV and 2962.0--2975.5 eV for \ce{[Rh(H2O)6]^{3+}}, and 12769.9--12784.0 eV and 11136.9--11147.3 eV for \ce{[Ir(H2O)6]^{3+}}. Here, the frequency windows used for each calculation were 0.61 eV (0.0225 Hartree), with steps of 0.068 eV (0.0025 Hartree). For the XAS calculations, channel restrictions\cite{STENER2003115,Debeer_2008,Besley2007,fransson2016} were imposed, where only transitions from the 2p orbitals (p$_{\frac{1}{2}}$ for the L$_{II}$-edge and p$_{\frac{3}{2}}$ for the L$_{III}$-edge) were allowed within the frequency windows.
From the resulting polarizabilities, $\alpha_{\alpha\beta}$ ($\alpha,\beta = x,yz,z$-coordinates), the cross-section $\sigma (\omega)$ was calculated as
\begin{equation}
	\label{eq:cross_section}
	\sigma (\omega)= \frac{4\pi \omega}{3c}\text{Im}(\alpha_{xx}+\alpha_{yy}+\alpha_{zz}) . 
\end{equation}
Here $\omega$ is the frequency of the corresponding polarizabilities, and $c$ is the vacuum speed of light.
The QM region in PE-CPP calculations was always chosen as \ce{[M(H2O)6]^{3+}} (M = Rh, Ir) and \ce{[Pt(N3)(OH)2(NH3)2]}. We employed the CAM-B3LYP\cite{Takeshi2004} functional with a def2-SV(P) basis set for the ligands, and a dyall.v2z\cite{dyall2009} basis set for the metals. All calculations were carried out using uncontracted basis sets. We used a non-colinear form of the exchange-correlation kernel as implemented in DIRAC.\cite{bast2009b} A damping factor ($\gamma$) of 4.557$\times 10^{-3}$ Hartree (1000 cm$^{-1}$) was used in all calculations. Relativistic effects were included through the use of both 4c and X2C Hamiltonians. The 4c calculations were carried out with (SS$|$SS) integrals replaced by interatomic SS energy corrections.\cite{visscher1997b} For the X2C calculations, atomic-mean-field calculations were used to obtain a correction for the two-electron spin-orbit interaction\cite{hess1996,AMFI1999}. To probe the quality of the atomic-mean-field corrections, we compared X2C with these corrections to full 4c-KS calculations for the aqua ions in vacuum and the resulting spectra are essentially identical (data not shown).  For \textit{trans}-Pt, only X2C calculations were carried out. To ensure that the calculated spectra did not contain spurious/artificial transitions our assignment of the transitions involved a careful examination of the response vectors. This included visual inspection of the molecular orbitals involved in transitions, as well as inspection of the atomic orbital contributions to the molecular orbitals in question.

\section{Results and Discussion \label{results}}
\subsection{Electronic structures of the model systems} 

Before discussing the UV-vis and X-ray spectra in the two subsections below, we first  make a few comments on the choice of test systems. We chose the aqua ions since they are among the simplest transition metal ions showing significant relativistic effects, allowing us to carry out both (PE-)4c and (PE-)X2C calculations without exhausting our computational resources. Further, they are inert $\text{d}^{6}$ aqua ions and therefore  well-defined in aqueous solution without fast exchange of water ligands; they have been experimentally characterized by UV-vis\cite{brorson1996,Gajhede1993} (in water) and in case of \ce{[Ir(H2O)6]^{3+}}, also by XAS.\cite{carrera2007} Further, their (approximately) octahedral symmetry dictates a d-orbital splitting with a full $t^{6}_{2g}$ set and empty $e_{g}$ orbitals. Thus, they can be expected to display types of excitations where the solvent effect is rather different, namely d-d transitions ($t_{2g}\rightarrow e_g$) and ligand-to-metal charge transfer (LMCT) transitions, involving the empty $e_g$ set of d-orbitals.  Analyzing the response vectors of the two aqua complexes for the transitions in the UV-vis  region (under 8 eV) showed indeed that all intense transitions involve empty metal d-orbitals (the occupied orbitals involved may either be ligand or other d-orbitals). For the \textit{trans}-Pt complex, only LMCT excitations are investigated and the nature of these transitions was found to be mainly from $\pi$ orbitals on the azide ligands (\ce{N3-}) to that of the d-orbitals on Pt. 

On one hand, the d-d transitions are not expected to be associated with large re-organization of the electron density during the excitation, and likely show little (electronic) effect of the solvent. On the other hand, LMCT transitions can be expected to show large re-organization of the electron density during excitation and hence larger (electronic) effect of the solvent is expected. We will investigate this in the following sections and we will in general separate solvent effects due to structural and electronic perturbation, although a few words of caution are needed: the dipole-forbidden d-d transitions can only gain intensity through symmetry lowering. This is partly possible in the water solvent, but for the vacuum-optimized structures, the only mechanism to lower the symmetry is through vibronic coupling. Vibrational effects are  not included in our calculations and a fair comparison to the vacuum-optimized structures is not possible in this case. We therefore only comment on electronic solvent contributions in this case. 	

Solvent effects including vibrational effects could be included through molecular dynamics (MDs) and this would indeed be required for a proper comparison with experimental results. However, we stress that our goal here is to quantify different solvent effects (and to investigate if the solvent effect is reproduced by X2C) rather than to make a detailed comparison with experiment. In this regard, it should also be noted that only the d-d transitions are experimentally observed for the aqua ions, since the LMCT transitions are too high in energy.  We include the LMCT in our analyses, although in some cases they may overlap with transitions for the water solvent (we discuss the consequences of this for the PE model below). The \textit{trans}-Pt complex will allow us to analyze a LMCT transition in the optically transparent window of the water solvent. 
	
We have also investigated the nature of electronic structure of the investigated complexes regarding relativistic effects: the d-orbitals in the two aqua complexes are heavily mixed in $\alpha$ or $\beta$-components, which is only correctly obtained if spin-orbit coupling is accounted for; see Tables S1 and S2 in the supporting information (SI). Our  analyses further showed the transitions at the L$_{\text{II}}$ and L$_{\text{III}}$-edges to be from occupied p$_{\frac{1}{2}}$ and p$_{\frac{3}{2}}$-orbitals to unoccupied d-orbitals (as expected) for \ce{[Rh(H2O)6]^{3+}} and \ce{[Ir(H2O)6]^{3+}}. Accordingly, the core-transitions are also metal-based and it could be expected that the solvent effects are similar (although differences are seen, as will be shown below). Still, in all cases the transitions in the aqua complexes cannot be described without the inclusion of spin-orbit coupling. We have previously investigated the platinum complex in the same manner and came to the same conclusion.\cite{creutzberg2020} We will in the subsections below therefore not discuss non-relativistic results, but focus on quantifying the effect of either solvation or the effect of switching from 4c to X2C Hamiltonian.

\subsection{UV-vis spectra of \ce{[Rh(H2O)6]^{3+}} and \ce{[Ir(H2O)6]^{3+}}}
\begin{figure}[htb!]
		\begin{subfigure}{0.49\textwidth}	
			\includegraphics[width=1.05\textwidth]{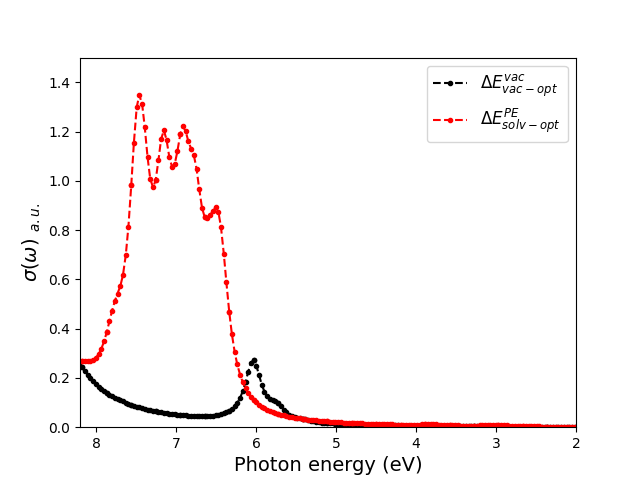}
			\caption{}
		\end{subfigure}
		\begin{subfigure}{0.49\textwidth}
			\includegraphics[width=1.05\textwidth]{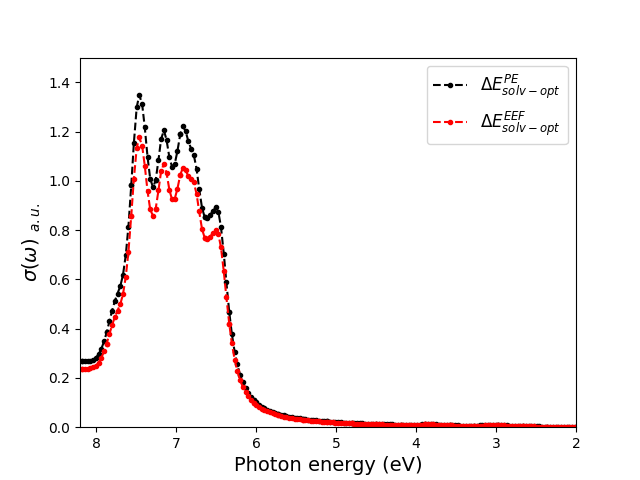}
			\caption{}
		\end{subfigure}
	\caption{UV-vis absorption spectra for \ce{[Rh(H2O)6]^{3+}}. (a) Comparison of spectra calculated in vacuum and with environment described through PE, where (b) compares spectra calculated in environment with PE and EEF effects.}
	\label{fig:uv-vis-Rh}
\end{figure}

\begin{figure}[htb!]
		\begin{subfigure}[b]{0.49\textwidth}	
			\includegraphics[width=1.05\textwidth]{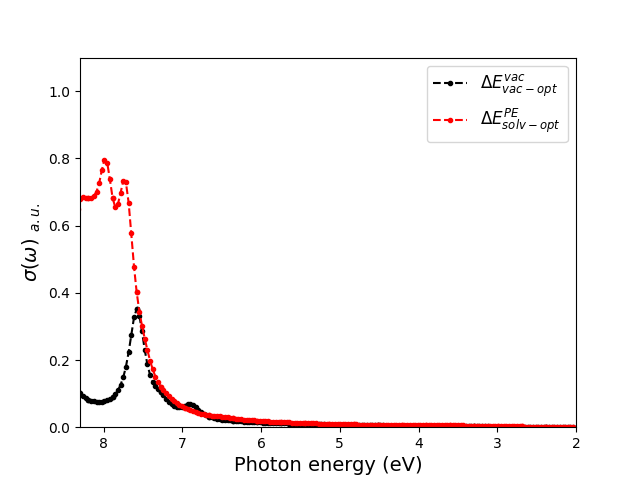} 
			\caption{}
		\end{subfigure}
		\begin{subfigure}[b]{0.49\textwidth}
			\includegraphics[width=1.05\textwidth]{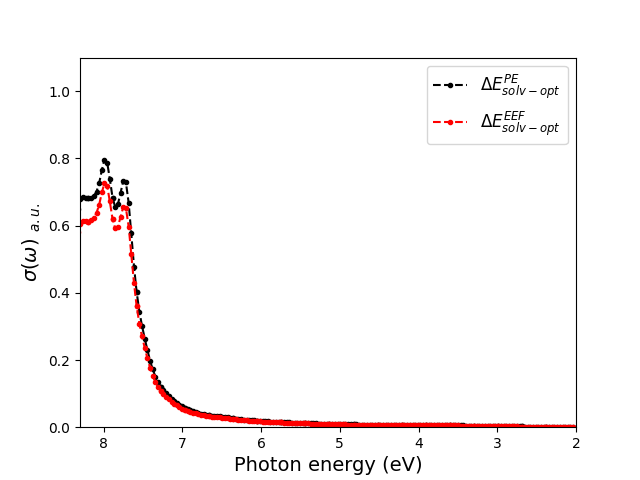}
			\caption{}
		\end{subfigure}
	\caption{UV-vis absorption spectra for \ce{[Ir(H2O)6]^{3+}}. (a) Comparison of spectra calculated in vacuum (including vacuum optimization) and with environment described through PE, where (b) compares spectra calculated in environment with PE and EEF effects.}
	\label{fig:uv-vis-Ir}
\end{figure}

The impact of going from vacuum to an aqueous environment on the UV-vis spectra of the two aqua complexes is shown in Figures \ref{fig:uv-vis-Rh} and \ref{fig:uv-vis-Ir}. We show the effect of PE in panel (a) and that of additionally including EEF effects in panels (b). Selected peak maxima and associated absorption cross-sections, $\sigma(\omega)$, are given in Tables \ref{tab:Rh-pe-uvvis} and \ref{tab:Ir-pe-uvvis}. Note that we in Figures  \ref{fig:uv-vis-Rh} and \ref{fig:uv-vis-Ir} only show comparison to vacuum-optimized structures while the Tables also includes calculations where we employ the structures optimized within the solvent, but without a PE potential.  The corresponding Figures comparing PE and vacuum calculations with solvent optimized structures are given in the SI (Figures S1 and S2).

As expected, the two complexes display intense ligand-to-metal charge transfer excitations at high energies and two lower-lying d-d transitions. The d-d transitions are not very intense and can not be discerned in Figures \ref{fig:uv-vis-Rh} and \ref{fig:uv-vis-Ir} (a magnified figure is provided in the SI, Figure S1 and S2).  Although our results for the d-d transitions correspond  well with experiment\cite{Gajhede1993,brorson1996} at 3.1 and 3.9 eV for \ce{[Rh(H2O)6]^{3+}} and 3.9 and 4.6--4.7 eV for \ce{[Ir(H2O)6]^{3+}} (see Tables \ref{tab:Rh-pe-uvvis} and \ref{tab:Ir-pe-uvvis}) we will not discuss them in detail, but rather focus on the LMCT transitions.  These transitions clearly demonstrate impact of solvation leads to qualitatively different spectra, where both excitation energies at peak maxima change and the absorption cross sections increase drastically. 

\begin{table}[htb!]
\centering
\caption{Selected peak maxima positions $\Delta E$  (in eV)  for \ce{[Rh(H2O)6]$^{3+}$} (the corresponding spectra are given in Figure \ref{fig:uv-vis-Rh}). The absorption cross sections, $\sigma(\omega)$, are given in parantheses.\label{tab:Rh-pe-uvvis}. For $\Delta E^{vac}_{vac-opt}$, the lack of vibronic coupling means that the d-d transitions have no intensity and they are therefore not reported.}
\begin{tabular}{llllc}
\hline
\hline \\[-2.0ex]
$\Delta E^{vac}_{vac-opt}$  & $\Delta E^{vac}_{solv-opt}$   & $\Delta E^{PE}_{solv-opt}$  & $\Delta E^{EEF}_{solv-opt}$ & Assignment \\[0.5ex] 
\hline \\[-1.5ex]
  -      &  7.63 (0.880)                &  7.46 (1.348) & 7.46 (1.176) & LMCT  \\[0.5ex] 
6.03 (0.272) &  7.08 (1.040)                &  7.15 (1.205) & 7.15 (1.070) & LMCT  \\[0.5ex]  
5.76 (0.107)&  6.81 (1.190)                &  6.91 (1.221) & 6.91 (1.053) & LMCT  \\[0.5ex] 
  - &  6.54 (0.952,  6.51 (0.948)  &  6.51 (0.893) & 6.51 (0.801) & LMCT  \\[0.5ex] 
  -          &  3.89 (0.015), 3.85 (0.015)  &  3.82 (0.013) & 3.82 (0.012) & d-d  \\[0.5ex] 
  -          &  3.14 (0.010)                &  3.11 (0.009) & 3.11 (0.008) & d-d  \\[0.5ex] 
  -          &  2.97 (0.010)                &  2.93 (0.009) & 2.93 (0.008) & d-d  \\[0.5ex] 
\hline
\end{tabular}
\end{table}

For the case with structures optimized in solvent (i.e.~the electronic solvent contributions), the peak maxima positions and their associated absorption cross-section are given in the $\Delta E^{vac}_{solv-opt}$ column in Tables \ref{tab:Rh-pe-uvvis} and \ref{tab:Ir-pe-uvvis}. 
For the LMCT transitions, the pure solvent contributions to the peak maxima vary between 0.0--0.17 eV for \ce{[Rh(H2O)6]^{3+}} and 0.07--0.17 eV for  \ce{[Ir(H2O)6]^{3+}}, showing that the PE contributions are non-negligible.

The absorption cross-sections are also significantly affected: for \ce{[Rh(H2O)6]^{3+}} at the peak maximum at 7.46 eV (with PE), we see an increase in $\sigma(\omega)$ from 0.880 a.u.~without PE to 1.348 a.u.~with PE, corresponding to an 53\% increase. For the other LMCT peaks, the corresponding increases are 16\% and 2\% for the peaks at 7.15 and 6.91 eV (with PE) or a decrease of 6\% for the peaks at 6.54 eV (with PE).  

For \ce{[Ir(H2O)6]^{3+}}, $\sigma(\omega)$ generally increases 20--27\% due to the PE environment: for the transition at 7.95 (with PE), $\sigma(\omega)$ increases from 0.619 a.u.~to 0.787 a.u.~(27\%), while  $\sigma(\omega)$ for the transition at 7.75 eV (with PE) increases from 0.610 a.u.~to 0.733 a.u.~(20 \%). 

The difference between d-d and LMCT transitions are clearly seen in their  response to the (pure) solvent effects (although we again emphasize that vibronic coupling is not accounted for). The d-d transitions show little effect of the solvent after the structural solvent perturbation is accounted for, whereas the LMCT excitations are associated with a much more pronounced electronic  solvent effect (see $\Delta E^{vac}_{solv-opt}$ and  $\Delta E^{PE}_{solv-opt}$ in Tables \ref{tab:Rh-pe-uvvis} and \ref{tab:Ir-pe-uvvis} as well as Figures S1 and S2).

We also investigated the effect of an EEF, which is shown on Figures \ref{fig:uv-vis-Rh}b and \ref{fig:uv-vis-Ir}b for \ce{[Rh(H2O)6]^{3+}} and \ce{[Ir(H2O)6]^{3+}}, respectively. As expected, the excitation energies at peak maxima positions are identical, but there is occasionally an effect on the absorption cross sections: to quantify the effect, we compare $\sigma(\omega)$ for the peak maxima positions in the $\Delta E^{PE}_{solv-opt}$ column with the corresponding numbers in the $\Delta E^{EEF}_{solv-opt}$ column in  Tables \ref{tab:Rh-pe-uvvis} and \ref{tab:Ir-pe-uvvis}. This comparison shows that $\sigma(\omega)$ for LMCT transitions are between 13-16 \% (\ce{[Rh(H2O)6]^{3+}})  and 10--12 \% (\ce{[Ir(H2O)6]^{3+}}) larger without EEF. For both complexes, the effect from the EEF is much smaller for the d-d transitions.
\begin{table}[htb!]
	\centering
	\caption{Selected peak maxima positions $\Delta E$  (in eV)  for \ce{[Ir(H2O)6]$^{3+}$} (the corresponding spectra are given in Figure \ref{fig:uv-vis-Ir}). The absorption cross sections, $\sigma(\omega)$, are given in parantheses. For $\Delta E^{vac}_{vac-opt}$, the lack of vibronic coupling means that the d-d transitions have no intensity and they are therefore not reported. \label{tab:Ir-pe-uvvis}}
	\begin{tabular}{lcccc}
		\hline
		\hline \\[-2.0ex]
		$\Delta E^{vac}_{vac-opt}$  & $\Delta E^{vac}_{solv-opt}$ & $\Delta E^{PE}_{solv-opt}$   & $\Delta E^{EEF}_{solv-opt}$   & Assignment \\[0.5ex] 
		\hline \\[-1.5ex]
		7.58 (0.351) & 7.79 (0.619)                & 7.95 (0.787) & 7.95 (0.717) & LMCT \\[0.5ex]
		6.90 (0.070) & 7.82 (0.610)                & 7.75 (0.733) & 7.75 (0.656) & LMCT \\[0.5ex]
		 & 7.58 (0.593)                & 7.72 (0.729) & 7.72 (0.652) & LMCT \\[0.5ex]
		-            & 4.25 (0.008), 4.21 (0.008)  & 4.18 (0.008) & 4.18 (0.007) & d-d  \\[0.5ex]
		-            & 3.77 (0.008), 3.74 (0.008)  & 3.70 (0.007) & 3.70 (0.007) & d-d  \\[0.5ex]
		\hline
	\end{tabular}
\end{table}

With respect to the LMCT excitations, it should be noted that they in some cases for \ce{[Ir(H2O)6]^{3+}} and for the highest-lying LMCT excitation for \ce{[Rh(H2O)6]^{3+}}  are close to resonant regions of the water solvent: the first excitation of water is around 7.5 eV\cite{mercedes2008}. The use of static polarizabilities will most likely lead to some error in these regions, but we leave investigation of this for further studies. A remedy may be to include parts of the solvent molecules fully in the QM region. We have so far not investigated QM size effects, since we present both four- and two-component calculations and the former becomes  computationally heavy with extended QM regions. As we will show in a section below, PE-4c and PE-X2C calculations provide close to identical description of the solvent effects, which is a good foundation for using X2C calculations in future investigations of the QM size. Alternatively, it is possible to employ  frequency-dependent polarizabilities, as has been done recently by several different groups\cite{harczuk2015,norby2017b,wildman2019} (within non-relativistic frameworks).  Still, frequency-dependent polarizabilities are not without challenges for response-based methods, although it has been pointed out\cite{norby2017b} that CPP methods are beneficial in this regard, since the frequency is known at the input stage. Again, we will not pursue this further in this paper, but rather turn to a LMCT excitation far from the resonant region of water, namely the one in the \textit{trans}-Pt complex investigated in next subsection.

\subsection{UV-vis spectrum of \textit{trans}-\textit{trans}-\textit{trans}-\ce{[Pt(N3)(OH)2(NH3)2]}}
\begin{figure}[htb!]
	\begin{subfigure}{0.49\textwidth}	
		\includegraphics[width=1.05\textwidth]{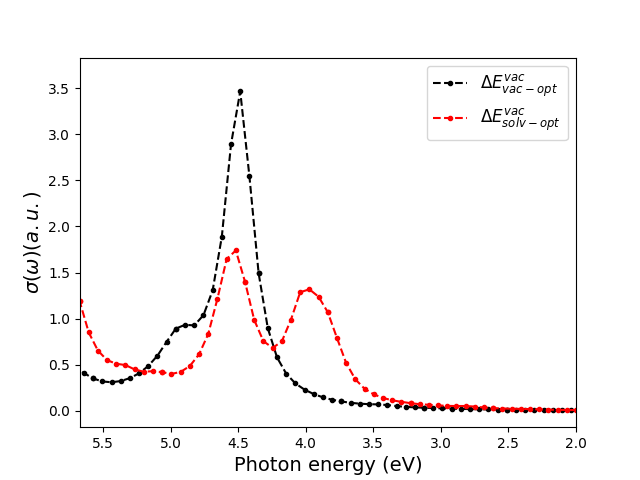}
		\caption{}
	\end{subfigure}
	\begin{subfigure}{0.49\textwidth}	
		\includegraphics[width=1.05\textwidth]{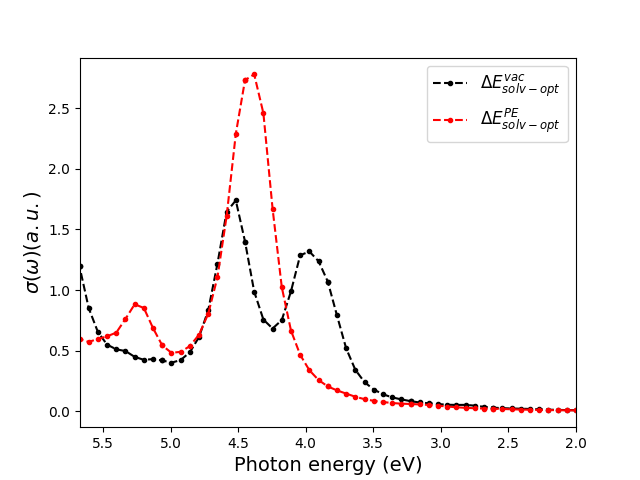}
		\caption{}
	\end{subfigure}
	\begin{subfigure}{0.49\textwidth}
		\includegraphics[width=1.05\textwidth]{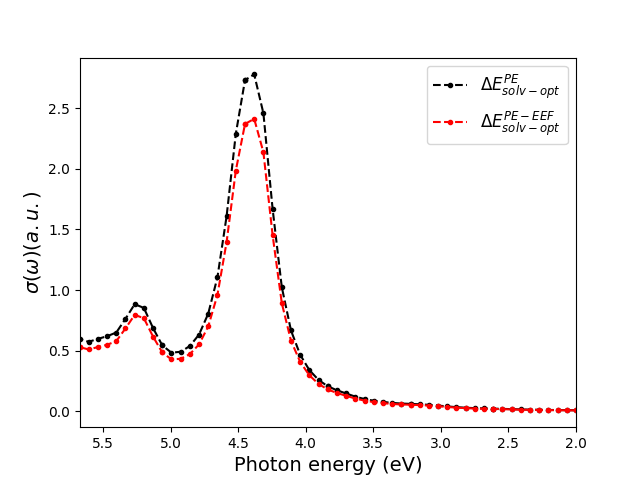}
		\caption{}
	\end{subfigure}
	\begin{subfigure}{0.49\textwidth}
		\includegraphics[width=1.05\textwidth]{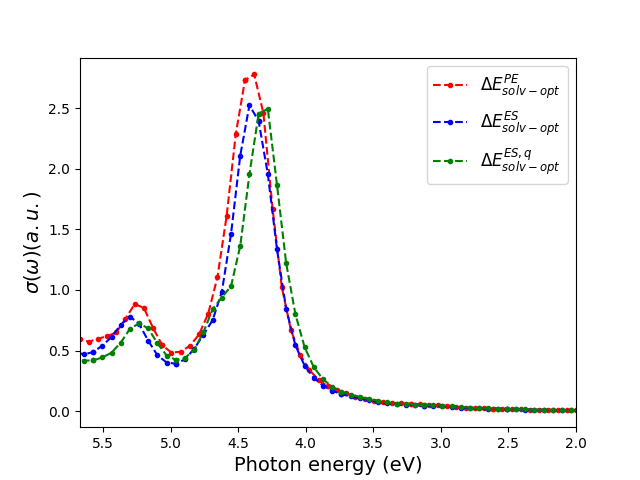}
		\caption{}
	\end{subfigure}
	\caption{UV-vis absorption spectra  for \textit{trans}-\textit{trans}-\textit{trans}-\ce{[Pt(N3)2(OH)2(NH3)2]}.(a) Comparison of spectra calculated in vacuum (including vacuum optimization) and with its environment described through PE. (b) Compares spectra calculated in vacuum (with solvent optimized structure) to that of spectra with PE. (c) Compares the effect of PE with an EEF to that of only using PE. (d) Compares the effect of PE to that of only including electrostatic embedding with either multipoles up to quadrupoles (ES)  or charges (ES,q).}
	\label{fig:uv-vis-Pt}
\end{figure}

The UV-vis spectra for \textit{trans}-Pt is shown in Figure \ref{fig:uv-vis-Pt}. We have previously calculated this spectrum in vacuum  with conventional response theory.\cite{creutzberg2020} We found that around 80 roots in the response equations were necessary to cover the UV-vis region; while this is possible for present days response solvers, it is certainly still high. Meanwhile, the PE-X2C-CPP is not affected by the high density of states: in  Figure \ref{fig:uv-vis-Pt} we show UV-vis spectra for (a) the structure optimization in vacuum and (b) with the structure optimized in solvent. From the figure, we see that including the solvent has a significant impact on the resulting spectra: the vacuum-optimized structure only shows one peak at 4.48 eV (with a shoulder at 4.89 eV), whereas the solvent-optimized structures has two intense peaks in the energy range 4.0--4.5 eV (experimentally\cite{mackay2006}, the complex has one large transition at 4.35 eV in water). Interestingly, the single intense peak is restored (now at 4.38 eV), while an additional (small) peak appears at 5.27 eV when a PE potential is included. The absorption cross section also changes significantly, as seen from Table \ref{tab:Pt-pe-uvvis}. Depending on whether we compare $\Delta E^{vac}_{vac-opt}$ or  $\Delta E^{vac}_{solv-opt}$ to $\Delta E^{PE}_{solv-opt}$ (see Table \ref{tab:Pt-pe-uvvis}), the absorption cross section changes from 3.464 a.u. or 1.319 a.u. to 2.779 a.u., corresponding to a decrease of 62 \% or an increase of 52 \% (the latter purely due to the electronic effect of the solvent). 

As in previous section, we also quantify the effect on an EEF, which is shown  in panel (c) of Figure \ref{fig:uv-vis-Pt}: while the effect is smaller than the effect of moving from vacuum to PE, this effect is not negligible; $\sigma(\omega)$ is around 12 \% larger without EEF.
\begin{table}[htb!]
	\centering
	\caption{Selected peak maxima positions $\Delta E$  (in eV)  for \textit{trans}-\textit{trans}-\textit{trans}-\ce{[Pt(N3)2(OH)2(NH3)2]}  (the corresponding spectra are given in Figure \ref{fig:uv-vis-Pt}. The absorption cross sections, $\sigma(\omega)$, are given in parantheses (in a.u.)\label{tab:Pt-pe-uvvis}. }
	\begin{tabular}{llllc}
		\hline
		\hline \\[-2.0ex]
		$\Delta E^{vac}_{vac-opt}$  & $\Delta E^{vac}_{solv-opt}$   & $\Delta E^{PE}_{solv-opt}$  & $\Delta E^{EEF}_{solv-opt}$ & Assignment \\[0.5ex] 
		\hline \\[-1.5ex]
		4.89 (0.934)      &  4.52 (1.740)    & 5.27 (0.883)  & 5.27 (0.795) &  LMCT \\[0.5ex] 
		4.48 (3.464)      &  3.97 (1.319)    & 4.38 (2.779)  & 4.38 (2.495)   &  LMCT \\[0.5ex] 
		\hline
	\end{tabular}
\end{table}
\begin{table}[htb!]
	\centering
	\caption{Selected peak maxima positions $\Delta E$  (in eV)  for \textit{trans}-\textit{trans}-\textit{trans}-\ce{[Pt(N3)2(OH)2(NH3)2]} with increasing sophistication of the PE potential (the corresponding spectra are given in Figure \ref{fig:uv-vis-Pt}(d). The absorption cross sections, $\sigma(\omega)$, are given in parantheses (in a.u.).\label{tab:Pt-pe-uvvis-breakdown}}
	\begin{tabular}{lllc}
		\hline
		\hline \\[-2.0ex]
		$\Delta E^{es,q}_{solv-opt}$ & $\Delta E^{es}_{solv-opt}$    & $\Delta E^{PE}_{solv-opt}$  & Assignment \\[0.5ex] 
		\hline \\[-1.5ex]
		5.23 (0.725)        &  5.30 (0.781)      & 5.27 (0.883)   &  LMCT  \\[0.5ex] 
		4.28 (2.495)        &  4.42 (2.524)      &  4.38 (2.779)    &   LMCT \\[0.5ex] 
		\hline
	\end{tabular}
\end{table}

In addition to the dissection of structural and electron solvent effects, we have for the \textit{trans}-Pt complex also dissected the electronic effect from the PE potential into the effect from multipoles and the additional effect of polarization. Morevoer, we also test the case where only charges are included. The spectra are compared in Figure \ref{fig:uv-vis-Pt} (d) and peak maxima positions together with corresponding absorption cross sections are given in Table \ref{tab:Pt-pe-uvvis-breakdown}. We focus again on the most intense LMCT transiton around 4.3--4.4 eV. As seen from the Figure, the additional effect of polarization is rather small for the excitation energy; in fact the large qualitative charge from the vacuum spectrum in Figures \ref{fig:uv-vis-Pt} (a)--(b) are introduced already by including the charges, although the higher order multiples have some effect on the excitation energy: adding multipoles increases the excitation energy with 0.14 eV (the effect on $\sigma(\omega)$ is only 1\%). The effect on the excitation energy with respect to adding polarization is smaller; it amounts to a 0.04 eV decrease. Yet, the effect on $\sigma(\omega)$ is considerably larger as the absorption cross section increases from 2.524 a.u.~to 2.779 a.u.~corresponding to 10 \%.

\subsection{X-ray atomic spectroscopy for \ce{[Rh(H2O)6]^{3+}} and \ce{[Ir(H2O)6]^{3+}}}

Before discussing the XAS spectra, it should be emphasized that the calculation of L-edge spectra is met with a number of issues not seen in corresponding UV-vis spectra calculations. Although the CPP method (relativistic or non-relativistic) has a number of advantages, it also comes with disadvantages and PE-CPP inherits both. One disadvantage (inherent to all response methods) is the lack of core-polarization, while another disadvantage (specific to the choice of DFT) is the self-interaction error; both of these issues can lead to large shifts on the transitions compared with experiment\cite{norman2018} and we also observe this here. Several schemes to improve the absolute excitation energies have been proposed (see e.g.~the work by South et al. \cite{south2016}), but since we here only focus on the effect of the environment, these schemes were not employed.  Moreover, it is not uncommon in the calculation of  L-edge spectra to see spurious transitions between valence- and high-lying virtual orbitals.\cite{south2016,fransson2016} As described more thoroughly elsewhere\cite{cacelli1991,ekstrom2006,south2016}, this artifact is related to the employed local Gaussian basis sets, which are not appropriate for continuum or near-continuum states. Indeed, our analysis of the response vectors revealed a small number of such spurious transitions for a few of the higher-lying L-edge peaks. The issue is not guaranteed to be resolved with larger (and more diffuse) basis sets\cite{south2016} and this has therefore not been pursued. Previous investigations have sometimes chosen to focus exclusively on excitations below the ionization threshold\cite{ekstrom2006,south2016} and we will also do that here (we estimate the ionization threshold from the orbital energy of either p$_{\frac{1}{2}}$  or p$_{\frac{3}{2}}$).\cite{south2016} A pragmatic approach is to invoke channel-restrictions as a remedy to spurious  transitions\cite{STENER2003115,Debeer_2008,Besley2007,fransson2016} and we have also employed this strategy. 
\begin{figure}[hbt!]
\begin{multicols}{2}	
	\begin{subfigure}[b]{0.49\textwidth}
	\includegraphics[width=1.05\textwidth]{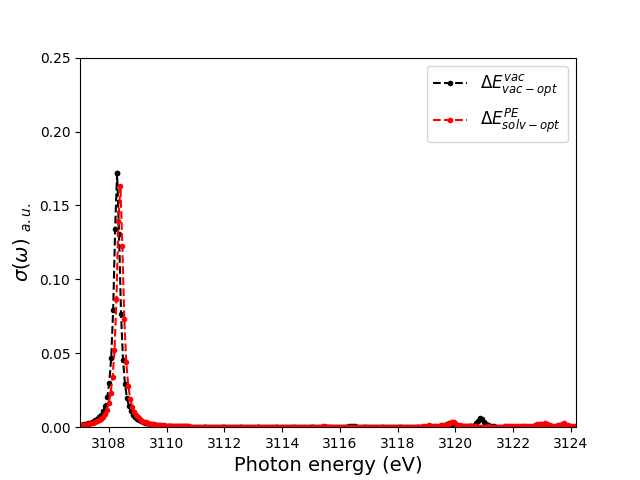}
    \caption{}
\end{subfigure}
\begin{subfigure}[b]{0.49\textwidth}
	\includegraphics[width=1.05\textwidth]{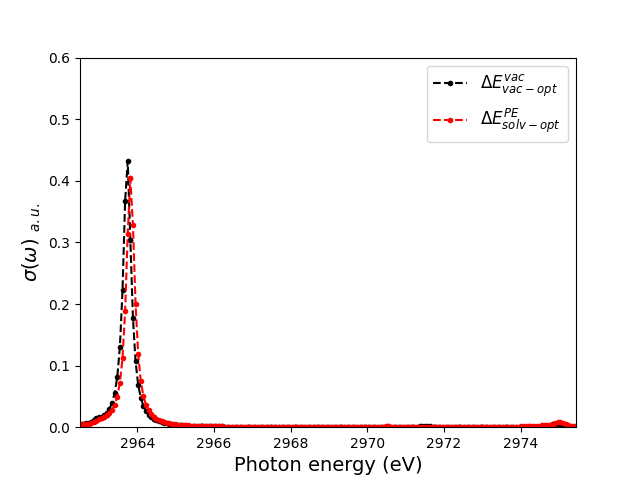}
	\caption{}
\end{subfigure}
    \begin{subfigure}[b]{0.49\textwidth} 
	\includegraphics[width=1.05\textwidth]{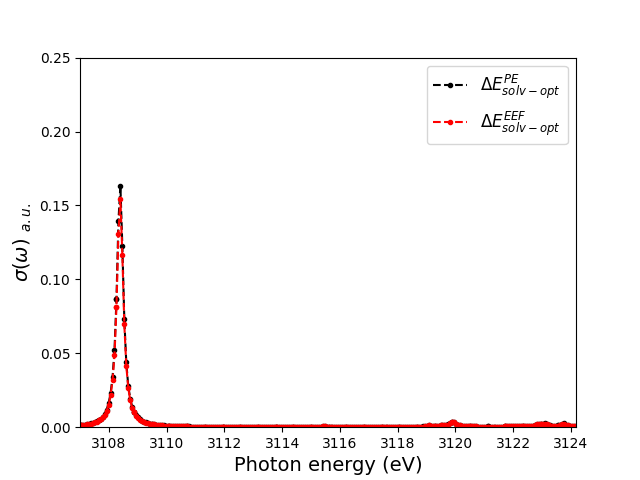}
	\caption{}
   \end{subfigure}
\begin{subfigure}[b]{0.49\textwidth}
	\includegraphics[width=1.05\textwidth]{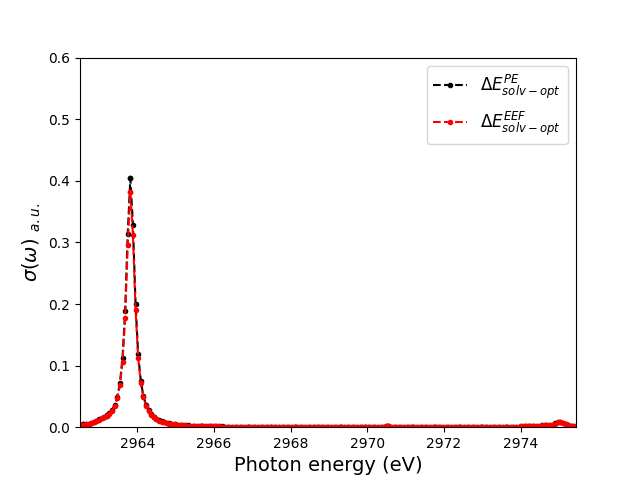}
    \caption{}
\end{subfigure}
\end{multicols}
	\caption{L-edge spectra of \ce{[Rh(H2O)6]^{3+}}. The panels (a) and (b) compares of vacuum and PE calculations for (a) L$_{\text{II}}$-edge and (b) the L$_{\text{III}}$-edge, while panels (c) and (d) compares PE calculations environment with and without EEF effects for (c) L$_{\text{II}}$-edge and (d) the L$_{\text{III}}$-edge. }    
	\label{fig:core_Rh}
\end{figure}
\begin{table}[htb!] 
	\centering
	\caption{Selected peak maxima positions $\Delta E$  (in eV)  for the L$_{\text{II}}$- and L$_{\text{III}}$-edges in \ce{[Rh(H2O)6]$^{3+}$} (the corresponding spectra are shown in Figure \ref{fig:core_Rh}). The absorption cross sections, $\sigma(\omega)$, are given in parantheses.  \label{tab:Rh-pe-core}}
	\begin{tabular}{lllll}
		\hline	
		\hline \\[-2.0ex]
		$\Delta E^{vac}_{vac-opt}$  & $\Delta E^{vac}_{solv-opt}$ & $\Delta E^{PE}_{solv-opt}$   & $\Delta E^{EEF}_{solv-opt}$   & Assignment \\[0.5ex] 
		\hline \\[-1.5ex]
		3108.28 (0.172)              & 3108.40 (0.171)             & 3108.40 (0.163)              & 3108.40 (0.155)                & p$_{\frac{1}{2}}$$\rightarrow$d (L$_{\text{II}}$)    \\[0.5ex]   
	2963.75  (0.432)           & 2963.81 (0.408)             &   2963.81  (0.404)           & 2963.81 (0.381)                & p$_{\frac{3}{2}}$$\rightarrow$d (L$_{\text{III}}$)   \\[0.5ex] 
		\hline
		\hline
	\end{tabular}
\end{table}
\begin{figure}[hbt!]
	\begin{multicols}{2}
		\begin{subfigure}[b]{0.49\textwidth}
			\includegraphics[width=1.05\textwidth]{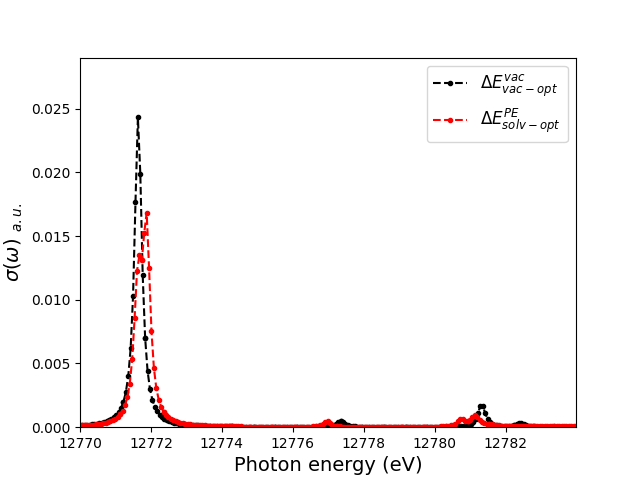}
			\caption{}
		\end{subfigure}
		\begin{subfigure}[b]{0.49\textwidth}
			\includegraphics[width=1.05\textwidth]{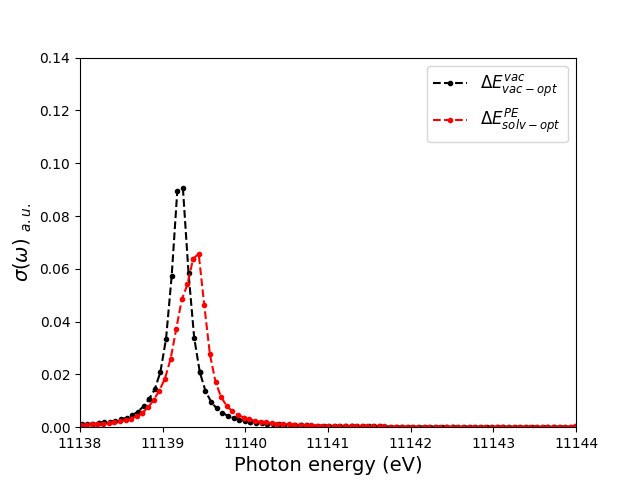}
			\caption{}
		\end{subfigure}
		\begin{subfigure}[b]{0.49\textwidth}
			\includegraphics[width=1.05\textwidth]{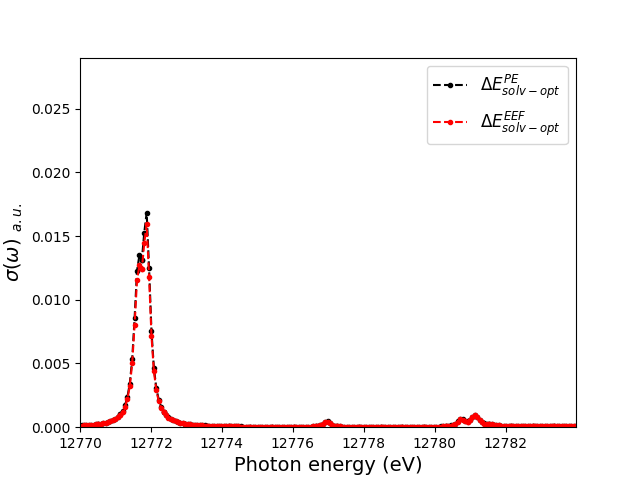}
			\caption{}
		\end{subfigure}
		\begin{subfigure}[b]{0.49\textwidth}
			\includegraphics[width=1.05\textwidth]{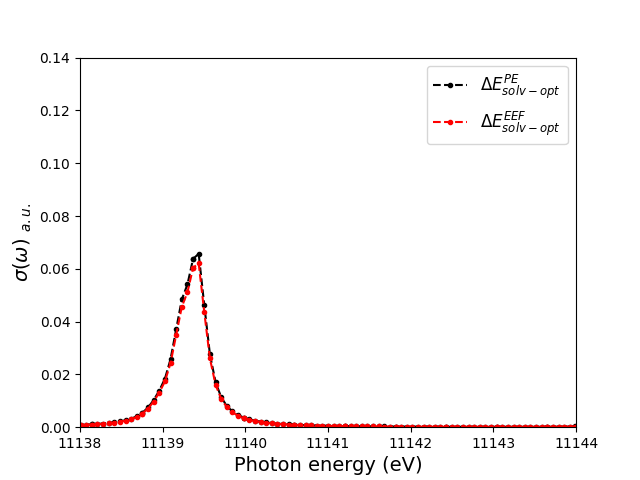}
			\caption{}
		\end{subfigure}
	\end{multicols}
	\caption{L-edge spectra of \ce{[Ir(H2O)6]^{3+}}. The panels (a) and (b) compares of vacuum and PE calculations for (a) L$_{\text{II}}$-edge and (b) the L$_{\text{III}}$-edge, while panels (c) and (d) compares PE calculations environment with and without EEF effects for (c) L$_{\text{II}}$-edge and (d) the L$_{\text{III}}$-edge.}    
	\label{fig:core_Ir}
\end{figure}
\begin{table}[htb!]
	\centering
	\caption{Selected peak maxima positions $\Delta E$  (in eV)  for the L$_{\text{II}}$- and L$_{\text{III}}$-edges in \ce{[Ir(H2O)6]$^{3+}$} (the corresponding spectra are shown in Figure \ref{fig:core_Ir}). The absorption cross sections, $\sigma(\omega)$, are given in parentheses. \label{tab:Ir-pe-core}}
	\begin{tabular}{lcccc}
		\hline
		\hline \\[-2.0ex]
		$\Delta E^{vac}_{vac-opt}$  & $\Delta E^{vac}_{solv-opt}$ & $\Delta E^{PE}_{solv-opt}$   & $\Delta E^{EEF}_{solv-opt}$   & Assignment \\[0.5ex] 
		\hline \\[-1.5ex]
		12771.6 (0.024) & 12771.8 (0.022) & 12771.9  (0.017)    & 12771.9 (0.016)    & p$_{\frac{1}{2}}$$\rightarrow$d (L$_{\text{II}}$)     \\[0.5ex]  
		 11139.2 ( 0.090)  &  11139.4 (0.085)   & 11139.4 (0.066) & 11139.4 (0.062)  & p$_{\frac{3}{2}}$$\rightarrow$d (L$_{\text{III}}$)    \\[0.5ex]
		\hline
	\end{tabular}
\end{table}
The L$_{\text{II}}$ and L$_{\text{III}}$ spectra are shown in Figures \ref{fig:core_Rh} and \ref{fig:core_Ir} while peak-maxima positions are provided in Tables \ref{tab:Rh-pe-core} and \ref{tab:Ir-pe-core} for \ce{[Rh(H2O)6]^{3+}} and \ce{[Ir(H2O)6]^{3+}}, respectively. As for the UV-vis spectra, the table also show results, $\Delta E^{vac}_{solv-opt}$, for spectra calculated in vacuum with underlying  solvent-optimized structures. The corresponding Figures are given in the SI (Figures S3 and S4). The calculated XAS spectra contain for both edges (and complexes) one main peak, corresponding to either a p$_{\frac{1}{2}}$$\rightarrow$d or a p$_{\frac{3}{2}}$$\rightarrow$d transition located on the metal. This transition corresponds to the main peak in the experimental spectrum\cite{carrera2007} of \ce{[Ir(H)2O)6]^{3+}}. As expected, the CPP value is shifted from the experiment, both due to the lack of orbital-polarization effects and self-interaction error in the DFT functional.\cite{south2016,fransson2016,norman2018} We note that two other peaks appear in the experimental spectrum at 16--17 eV and 44 eV above the main peak.\cite{carrera2007} None of these peaks are within the frequency window for our L$_{\text{III}}$-edge calculations, wheres we for the L$_{\text{II}}$-edge included up to around 12 eV above the main peak. While we do see small appearances of peaks around 5-10 eV above the main peak, we do not include these in our discussion as they are just around the ionization threshold (12782 eV for L$_{\text{II}}$ in the PE calculations). 
\begin{figure}[hbt!]
	\begin{multicols}{2}
		\begin{subfigure}[b]{0.49\textwidth}
			\includegraphics[width=1.05\textwidth]{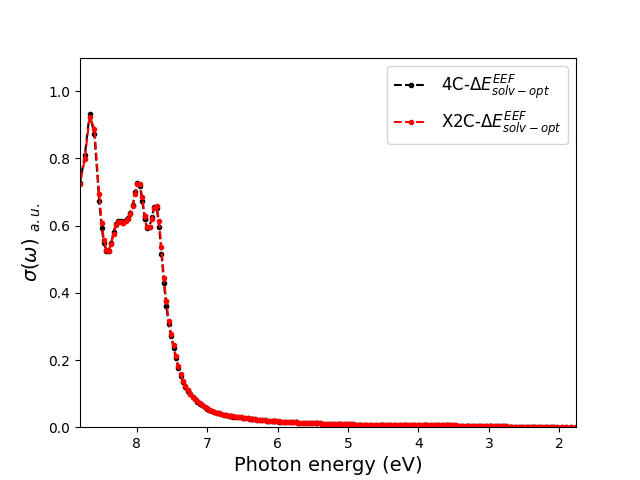}
			\caption{}
		\end{subfigure}
		\begin{subfigure}[b]{0.49\textwidth}
			\includegraphics[width=1.05\textwidth]{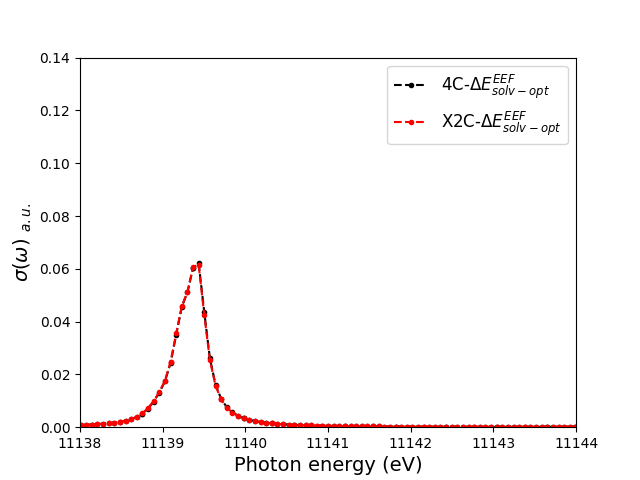}
			\caption{}
		\end{subfigure}
		\begin{subfigure}[b]{0.49\textwidth}
			\includegraphics[width=1.05\textwidth]{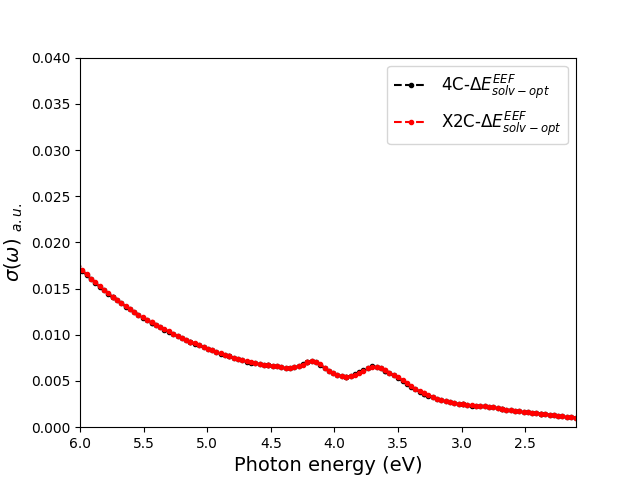}
			\caption{}
		\end{subfigure}
		\begin{subfigure}[b]{0.49\textwidth}
			\includegraphics[width=1.05\textwidth]{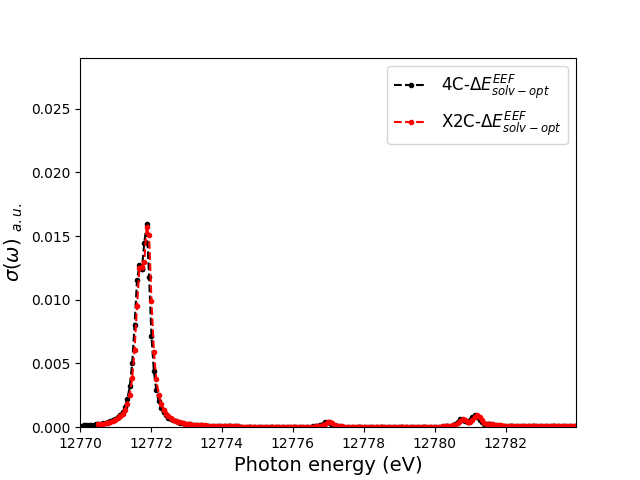}
			\caption{}
		\end{subfigure}
	\end{multicols}
	\caption{PE-4c- and PE-X2C-CPP spectra in UV-vis and XAS regions for \ce{[Ir(H2O)6]^{3+}}. (a) Full UV-vis spectrum and (c) focuses on the d-d transitions. The XAS spectra are shown in (b) for the L$_{\text{III}}$ (X2C spectra shifted 12.9 eV) and (d) L$_{\text{II}}$ (X2C spectra shifted 30.61 eV).}    
	\label{fig:Ir_4c_x2c_uvvis}
\end{figure}

Focusing now on the main transition, this excitation is located entirely on orbitals on the metal. In relation to any potential solvent effect, we therefore anticipated a case similar to d-d transitions discussed above, i.e., only benign effects from the solvent. For the \ce{[Rh(H2O)6]^{3+}} complex, this is indeed the case. However,  for the \ce{[Ir(H2O)6]^{3+}} complex, the change in the absorption cross-section is significant:  we obtain a decrease in $\sigma(\omega)$ from 0.024 a.u.~to 0.017 with PE and to 0.016 a.u.~with EEF for the L$_{\text{II}}$-edge. The corresponding numbers for the L$_{\text{III}}$-edge are decreases of 0.090 a.u.~to 0.066 a.u.~(PE) or 0.062 a.u.~(EEF), cf.~Table \ref{tab:Ir-pe-core}.  Thus, $\sigma(\omega)$ decreases 29-33 \% for the  L$_{\text{II}}$-edge and 27-31\% for the L$_{\text{III}}$-edge. By comparing calculations without PE in the solvated structure ($\Delta E^{vac}_{solv-opt}$ in  Table \ref{tab:Ir-pe-core}) to the pure vacuum calculation ($\Delta E^{vac}_{vac-opt}$), we see that the structural effect is rather small for the absorption cross section. Thus, the change is rather introduced by the electronic (PE) part, $\Delta E^{PE}_{solv-opt}$ in Table \ref{tab:Ir-pe-core}, showing that the environment electrostatics can  be important, also for L-edge XAS spectroscopy. Note also that EEF effects only account for 11-12 \% of the total change.

\subsection{Comparing four- and exact two-component calculations}

In this last section, we investigate the impact of using a two-component Hamiltonian (X2C) on the solvated UV-vis and XAS spectra. This is done by comparing the solvated spectra calculated with PE-X2C-CPP to their PE-4c-CPP counterparts. The results are shown in Figures \ref{fig:Ir_4c_x2c_uvvis} for \ce{[Ir(H2O)6]^{3+}} (results for \ce{[Rh(H2O)6]^{3+}} are similar and provided in the SI, Figure S5). The UV-vis spectra are practically identical, whereas the change of Hamiltonian leads to shifts in absolute excitation energies in the core spectra. However, applying this shift as done in Figure \ref{fig:Ir_4c_x2c_uvvis}, the spectra become identical, showing the PE-X2C-CPP method works extremely well.

\section{Conclusion}
We have presented the derivation and implementation of the CPP method in a relativistic  PE framework. The implementation is illustrated by calculations of both UV-vis and XAS spectra of solvated aqua-complexes \ce{[Rh(H2O)6]^{3+}} and  \ce{[Ir(H2O)6]^{3+}}. We next move to the larger complex \textit{trans}-\textit{trans}-\textit{trans}-\ce{[Pt(N3)2(OH)2(NH3)2]}, where we also show that inclusion of the environment can qualitatively change the obtained spectra.  In this case, we show that the largest electronic effect is from the charges, but polarizabilities in the environment have non-negligible effect on the absorption cross section. 

In all cases, the complexes required a relativistic method to correctly capture their electronic structures, while treatment of the environment is indispensable for a correct physical description of the UV-vis and XAS spectra. Finally, our calculations show that a PE-X2C-CPP model essentially reproduces corresponding  PE-4c-CPP spectra (this was done for the aqua ions); this highlights that the solvent effects are described well on the two-component level.

While our current target systems are experimentally known, we mostly consider them toy models.  We are currently investigating the dynamic effects of the solvent, which is not included in our present calculations. This is mainly due to the large computational demands for these types of calculations and the challenge with lack of  accurate force field parameters for transition metal systems. Another line of work we will pursue is to extend the method to electronic circular dichroism (ECD) spectroscopy.

\begin{acknowledgement}

The authors thank The Villum Foundation, Young Investigator Program (grant no. 29412), the Swedish Research Council (grant no. 2019-04205), and Independent Research Fund Denmark (grant no. 0252-00002B) for support. The computations were performed on computer resources provided by the Swedish National Infrastructure for Computing (SNIC) at Lunarc (Lund University) and HPC2N (Ume\aa\  University).

\end{acknowledgement}


\begin{suppinfo}

The supporting information contains Tables with Mulliken ($\alpha$ and $\beta$) populations  of the metal d-orbitals and energies of metal 2p orbitals. Moreover, the SI contains spectra for the solvent optimized hexaaqua ions without PE and a zoom in of the region with d-d transitions.

\end{suppinfo}

\bibliography{biblio}

\end{document}


\begin{table}[H]
\begin{center}
\hspace*{-2.0cm}
\centering
\scalebox{0.8}{
\begin{tabular}{ |c c c |}
 \hline
 \hline
  \multicolumn{3}{c}{4c Spinors}       \\ 
 \hline
 \hline
 Orbital/spinor  & Character & $\alpha$ ; $\beta$ \\ 
 \hline
\hline
  49 & $Rh(d)$ &   0.7065 ; 0.2935  \\
  50 & $Rh(d)$ &   0.8547 ; 0.1453  \\
  51 & $Rh(d)$ &   0.2799 ; 0.7201  \\
  52 & $Rh(d)$ &   0.9912 ; 0.0088  \\
  53 & $Rh(d)$ &   0.9736 ; 0.0264  \\
  54 & $Rh(s)$ &   1.0000 ; 0.0000  \\
  \hline
\end{tabular}
}
\\[0.2ex]
\caption*{Table S1: $\alpha$ and $\beta$ occupation numbers for orbitals/spinors with d-character involved in the transitions for \ce{[Rh(H2O)6]^{3+}} (solvated with PE)  calculated with 4c-CAM-B3LYP.} 
\end{center}
\end{table}

\begin{table}[H]
	\begin{center}
		\hspace*{-2.0cm}
		\centering
		\scalebox{0.8}{
			\begin{tabular}{ |c c c |}
				\hline
				\hline
				\multicolumn{3}{c}{4c Spinors}       \\
				\hline
				\hline
				Orbital/spinor  & Character & $\alpha$ ; $\beta$ \\ 
				\hline
				\hline
				65 & $Ir(d)$ &  0.8350 ; 0.1650  \\
				66 & $Ir(d)$ &  0.7010 ; 0.2990  \\
				68 & $Ir(d)$ &  0.5207 ; 0.4793  \\
				69 & $Ir(d)$ &  0.7037 ; 0.2963  \\
				\hline
			\end{tabular}
		}
		\\[0.2ex]
		\caption*{Table S2: $\alpha$ and $\beta$ occupation numbers for orbitals/spinors with d-character involved in the transitions for \ce{[Ir(H2O)6]^{3+}} (solvated with PE) calculated with 4c-CAM-B3LYP.
		} 
	\end{center}
\end{table}

\begin{table}[htb!]
	\centering
	\begin{tabular}{lccc}
		\hline
		\hline \\[-2.0ex]
		Orbitals & $\Delta E^{vac}_{vac-opt}$  & $\Delta E^{vac}_{solv-opt}$ & $\Delta E^{PE}_{solv-opt}$    \\[0.5ex]
		\hline
		\\[-1.5ex]
		3    &  3131.79   & 3131.66    & 3119.63 
		\\[0.5ex]
		4    & 2986.85    & 2986.73    & 2974.70 
		\\[0.5ex]
		5   & 2986.85  & 2986.72    & 2974.69  
		\\[0.5ex]
		\hline
	\end{tabular}
	\caption*{Table S3: Energies of $2p$-orbitals in eV for \ce{[Rh(H2O)6]^{3+}} } 
\end{table}


\begin{table}[htb!]
	\centering
	\begin{tabular}{lccc}
		\hline
		\hline \\[-2.0ex]
		Orbitals & $\Delta E^{vac}_{vac-opt}$  & $\Delta E^{vac}_{solv-opt}$ & $\Delta E^{PE}_{solv-opt}$    \\[0.5ex]
		\hline
		\\[-1.5ex]
		3 (p$_{\frac{1}{2}}$)   & 12793.21   & 12793.23   & 12782.01 
		\\[0.5ex]
		4 (p$_{\frac{3}{2}}$)    & 11160.55    &  11160.57   & 11149.35
		\\[0.5ex]
		5 (p$_{\frac{3}{2}}$)   & 11160.55  & 11160.57   &  11149.35 
		\\[0.5ex]
		\hline
	\end{tabular}
	\caption*{Table S4: Energies of $2p$-orbitals in eV  for \ce{[Ir(H2O)6]^{3+}}.} 
\end{table}




\begin{figure}[htb!]
	\begin{multicols}{2}
		\begin{subfigure}[b]{0.49\textwidth}	
			\includegraphics[width=1.05\textwidth]{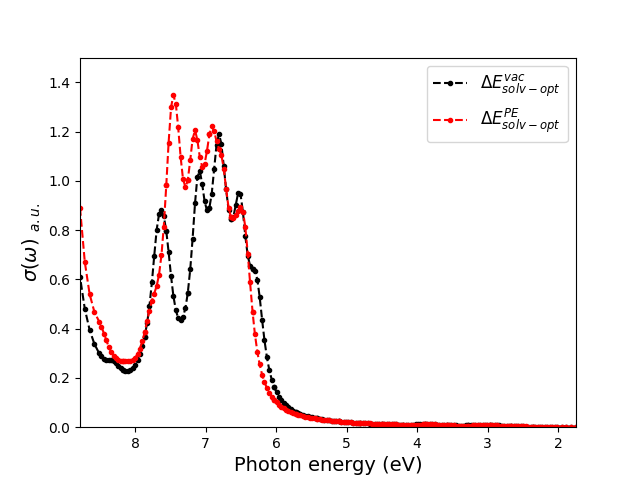}
			\caption{}
		\end{subfigure}
		\begin{subfigure}[b]{0.49\textwidth}
			\includegraphics[width=1.05\textwidth]{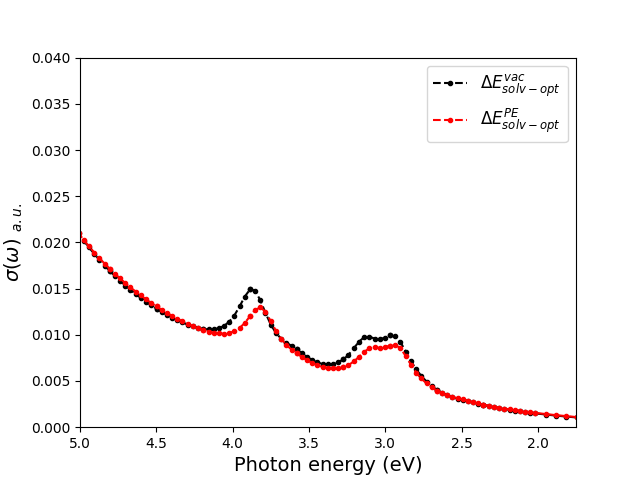}
			\caption{}
		\end{subfigure}
	\end{multicols}
	\caption*{Figure S1: 4c-CAM-B3LYP spectra with and without PE, employing the \ce{[Rh(H2O)6]^{3+}} structure QM/MM  optimized in solvent. (a) Full spectra and (b) focus on d-d region.}
\end{figure}

\begin{figure}[htb!]
	\begin{multicols}{2}
		\begin{subfigure}[b]{0.49\textwidth}	
			\includegraphics[width=1.05\textwidth]{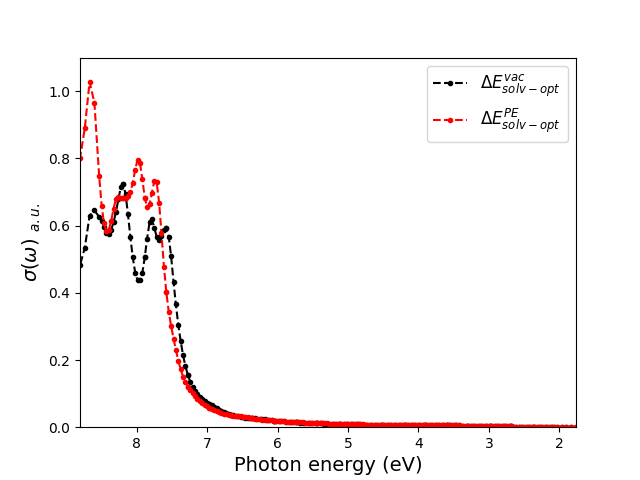}
			\caption{}   
		\end{subfigure}
		\begin{subfigure}[b]{0.49\textwidth}
			\includegraphics[width=1.05\textwidth]{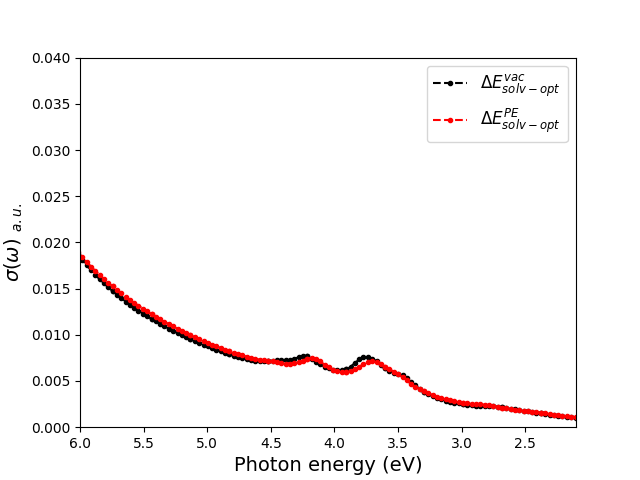}
			\caption{}
		\end{subfigure}
	\end{multicols}
	\caption*{Figure S2: 4c-CAM-B3LYP spectra with and without PE, employing the \ce{[Ir(H2O)6]^{3+}} structure QM/MM  optimized in solvent. (a) Full spectra and (b) focus on d-d region.}
\end{figure}


\begin{figure}[htb!]
	\begin{multicols}{2}
		\begin{subfigure}[b]{0.49\textwidth}	
			\includegraphics[width=1.05\textwidth]{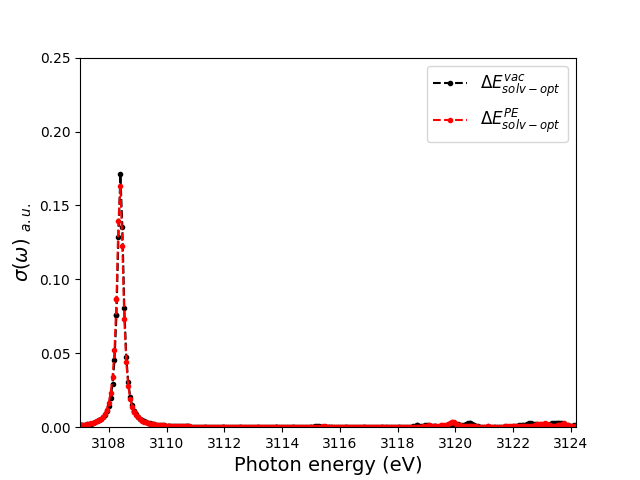}
			\caption{}
		\end{subfigure}
		\begin{subfigure}[b]{0.49\textwidth}
			\includegraphics[width=1.05\textwidth]{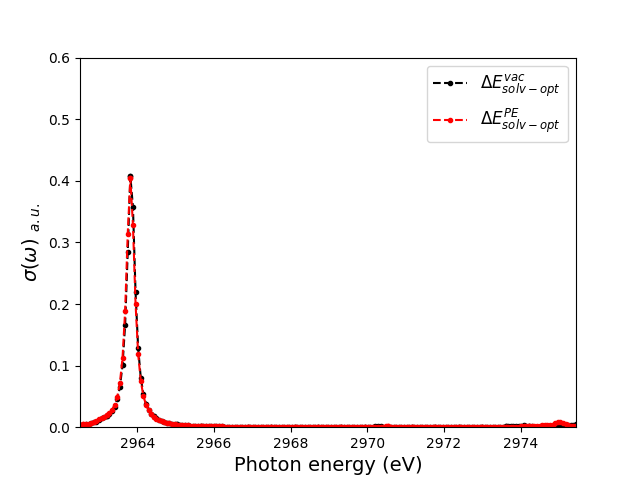}
			\caption{}
		\end{subfigure}
	\end{multicols}
	\caption*{Figure S3: 4c-CAM-B3LYP spectra with and without PE, employing the \ce{[Rh(H2O)6]^{3+}} structure QM/MM  optimized in solvent. (a) L$_{II}$ edge (b) L$_{III}$.}
\end{figure}

\begin{figure}[htb!]
	\begin{multicols}{2}
		\begin{subfigure}[b]{0.49\textwidth}	
			\includegraphics[width=1.05\textwidth]{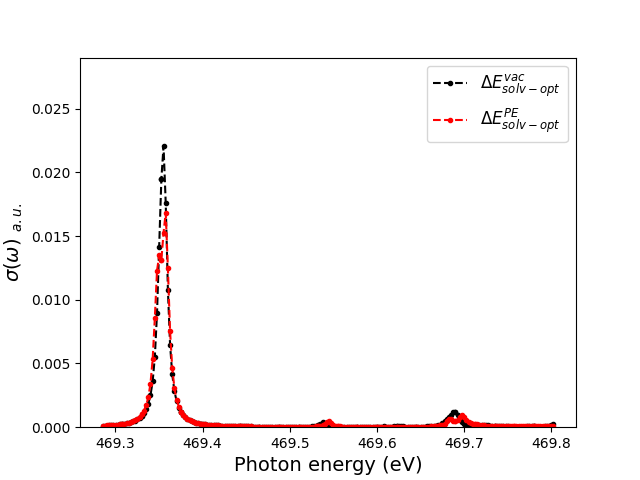}
			\caption{}
		\end{subfigure}
		\begin{subfigure}[b]{0.49\textwidth}
			\includegraphics[width=1.05\textwidth]{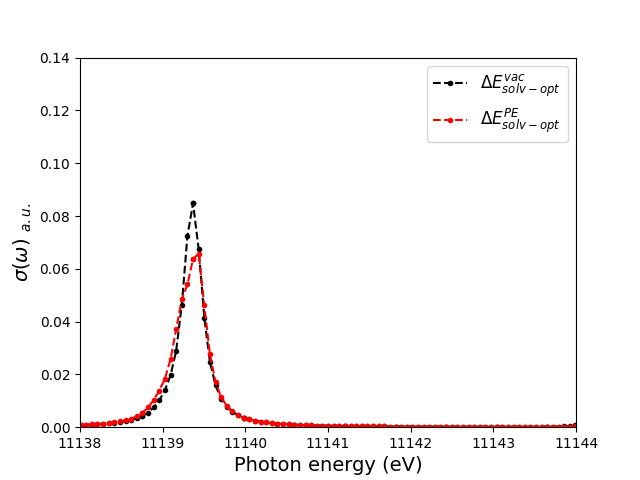}
			\caption{}
		\end{subfigure}
	\end{multicols}
	\caption*{Figure S4: 4c-CAM-B3LYP spectra with and without PE, employing the \ce{[Ir(H2O)6]^{3+}} structure QM/MM  optimized in solvent. (a) L$_{II}$ edge (b) L$_{III}$.}
\end{figure}

\begin{figure}[hbt!]
	\begin{multicols}{2}
		\begin{subfigure}[b]{0.49\textwidth}
			\includegraphics[width=1.05\textwidth]{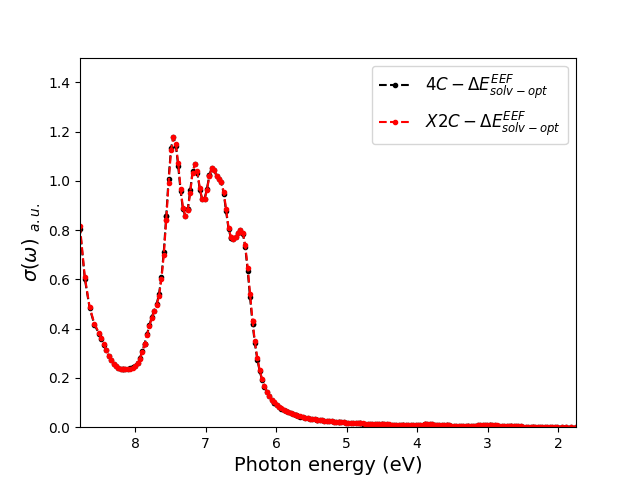}
			\caption{}
		\end{subfigure}
		\begin{subfigure}[b]{0.49\textwidth}
			\includegraphics[width=1.05\textwidth]{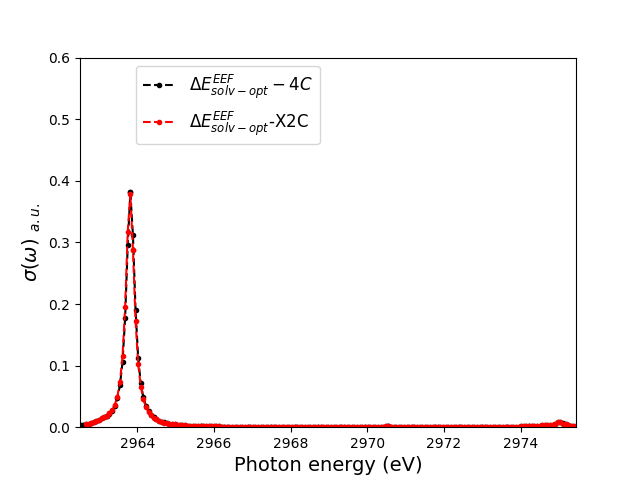}
			\caption{}
		\end{subfigure}
		\begin{subfigure}[b]{0.49\textwidth}
			\includegraphics[width=1.05\textwidth]{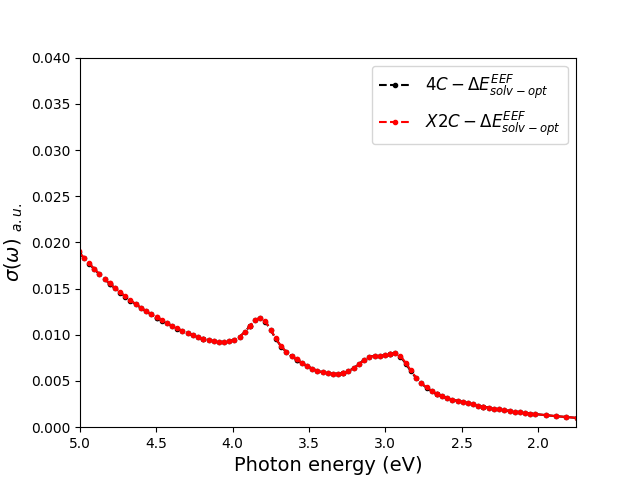}
			\caption{}
		\end{subfigure}
		\begin{subfigure}[b]{0.49\textwidth}
			\includegraphics[width=1.05\textwidth]{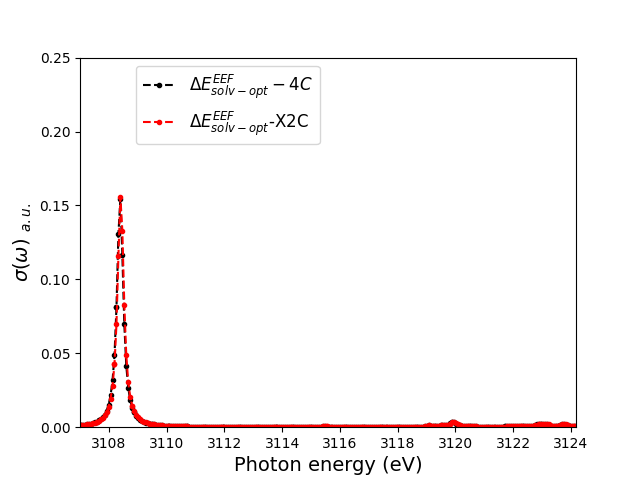}
			\caption{}
		\end{subfigure}
	\end{multicols}
	\caption*{Figure S5: PE-4c- and X2C spectra in UV-vis and XAS regions for \ce{[Rh(H2O)6]^{3+}}. (a) Full UV-vis spectrum and (c) zoom on the d-d transitions. The XAS spectra are shown in (b) for  L$_{III}$ (X2C spectra shifted  1.48 eV) and (d) L$_{II}$ (X2C spectra shifted 3.51 eV).}    
	\label{fig:Rh_4c_x2c_uvvis}
\end{figure}
